\documentclass[%
reprint,
superscriptaddress,
 amsmath,amssymb,
 aps,
 prl,
]{revtex4-2}

\usepackage{lipsum}

\usepackage{graphicx}%
\usepackage{dcolumn}%
\usepackage{bm}%

\usepackage[normalem]{ulem}

\usepackage{ragged2e}

\usepackage{xcolor}

\definecolor{fedeorange}{rgb}{0.98, 0.7, 0.2}

\definecolor{giocolor}{RGB}{0, 150, 100}

\usepackage{orcidlink}

\begin{document}

\preprint{APS/123-QED}

\title{Interplay between evolutionary and epidemic time scales challenges the outcome of control policies}

\author{Santiago Lamata-Ot\'in\,\orcidlink{0009-0004-0247-4792}}
\affiliation{Department of Condensed Matter Physics, University of Zaragoza, 50009 Zaragoza (Spain).}
\affiliation{GOTHAM lab, Institute for Biocomputation and Physics of Complex Systems (BIFI), University of Zaragoza, 50018 Zaragoza (Spain).}

\author{Alex Arenas\,\orcidlink{0000-0003-0937-0334}}
\affiliation{Departament d’Enginyeria Informàtica i Matemàtiques, Universitat Rovira i Virgili, 43007 Tarragona, Spain}
\affiliation{Pacific Northwest National Laboratory, 902 Battelle Boulevard, Richland, Washington 99354, USA}
\affiliation{Complexity Science Hub Vienna, Metternichgasse 8, 1030 Vienna, Austria}
\affiliation{ComSCIAM, Universitat Rovira i Virgili, 43007 Tarragona, Spain}

\author{Jes\'us G\'omez-Garde\~nes\,\orcidlink{0000-0001-5204-1937}}
\affiliation{Department of Condensed Matter Physics, University of Zaragoza, 50009 Zaragoza (Spain).}
\affiliation{GOTHAM lab, Institute for Biocomputation and Physics of Complex Systems (BIFI), University of Zaragoza, 50018 Zaragoza (Spain).}
\affiliation{Center for Computational Social Science, University of Kobe, 657-8501 Kobe (Japan).} 

\author{David Soriano-Pa\~nos\,\orcidlink{0000-0002-6388-4056}}%
\thanks{sorianopanos@gmail.com}
\affiliation{Departament d’Enginyeria Informàtica i Matemàtiques, Universitat Rovira i Virgili, 43007 Tarragona, Spain}
\affiliation{GOTHAM lab, Institute for Biocomputation and Physics of Complex Systems (BIFI), University of Zaragoza, 50018 Zaragoza (Spain).}

\date{\today}

\begin{abstract}

The classical SIR model, assuming constant viral traits, represents the cornerstone model for computing key indicators during epidemic outbreaks, such as the expected peak of infections or the impact of control policies. Viral evolution has been reported to challenge the physics of the SIR model, changing the nature of the epidemic transitions or the early-time dynamics of outbreaks. {Here we consider a minimal extension of the SIR model, allowing infectiousness to evolve,} to explore how the latter mechanism affects the two aforementioned indicators. We show that evolution induces a non-monotonic behavior of the epidemic peak with the basic reproduction number and undermines the impact of control policies, as lifting interventions too early can lead to worse epidemic scenarios than no action. We derive analytical expressions for the critical mutation rate and intervention time governing this behavior and identify a strong asymmetry between control strategies: while shortening the infectious period hinders transmission without suppressing the evolution of viral infectiousness, lowering transmission both reduces cases and slows down this evolution.

\end{abstract}

\maketitle

\noindent\textit{Introduction.—}
The dynamics of infectious diseases emerge from the interplay between pathogen traits and host populations \cite{keeling2008modeling,pastor2015epidemic}. Classical epidemic models, such as the susceptible–infected–recovered (SIR) framework \cite{kermack1927contribution,hethcote2000mathematics}, have been instrumental in understanding outbreak dynamics and in guiding the design of control strategies during recent pandemics \cite{flaxman2020estimating,hsiang2020effect,kissler2020projecting,arenas2020modeling}. Recent theoretical studies \cite{di2021optimal,bliman2021optimal,ketcheson2021optimal,perkins2020optimal,hindes2016epidemic,rozan2026analysis}, and particularly the work of Morris et al. \cite{morris2021optimal}, have further analyzed how simple, fixed-strength interventions can be robust and near-optimal control strategies.

A key underlying assumption in these approaches is that pathogen traits remain effectively constant over the course of an outbreak. Yet, many pathogens evolve on timescales comparable to epidemic spreading, so that mutation can directly affect epidemic dynamics \cite{grenfell2004unifying,anderson1991infectious,day2007applying,lion2018beyond,tsimring1996rna}. The impact of control policies on viral evolution has been mostly addressed in two-strain settings, characterizing analytically how interventions alter the emergence, survival and immune escape probabilities of mutants both within hosts~\cite{iwasa2003evolutionary,iwasa2004evolutionary,hartfield2015within,van2021controlling} and at the population level~\cite{hartfield2014epidemiological,meehan2020probability,ashby2023non}. Beyond the two-strain settings, phylodynamic approaches have proven substantial alterations in the evolutionary trajectories of Ebola and Influenza A viruses following non-pharmaceutical interventions applied to contain  Ebola~\cite{dellicour2018phylodynamic} or COVID-19 outbreaks respectively~\cite{chen2024covid,chen2025disruption}.

The theoretical study of how viral evolution shapes epidemic trajectories in multi-strain settings has been much less explored in the literature~\cite{zhang2022epidemic,soriano2025eco}. It has been reported that evolution induces super-exponential growth of outbreaks or abrupt epidemic transitions~\cite{zhang2022epidemic} in the SIR model; nonetheless, the impact of evolution on pivotal indicators such as the peak of infections or the impact of control policies remains a theoretical challenge. In this Letter, we address these questions {within a minimal extension of the SIR model in which infectivity is an evolving pathogen trait \cite{zhang2022epidemic}.} We show that evolution reverses the monotonic dependence of the epidemic peak on the basic reproduction number and the short-term benefits of epidemic control policies. We also find strongly asymmetric outcomes of interventions shaping within-host or inter-host dynamics under evolution. Our results therefore show that epidemic control and pathogen evolution are intrinsically coupled and should therefore be studied jointly for rapidly evolving viruses.
\medskip

\begin{figure}
    \centering
    \includegraphics[width=1\linewidth]{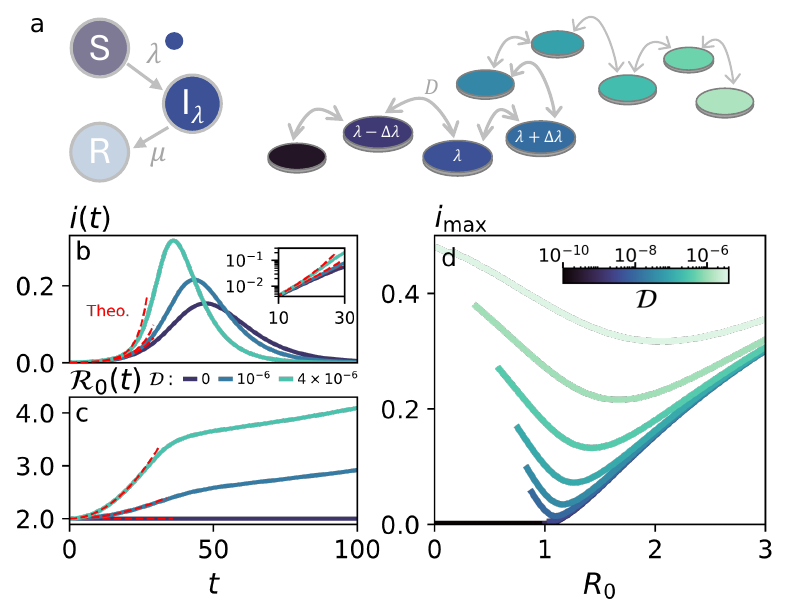}
    \caption{\textbf{Effect of infectivity evolution on epidemic trajectories.} \textbf{a} Schematic illustration of the epidemic model with evolving infectivity. \textbf{b-c} Temporal evolution of the epidemic prevalence $i(t)$ (b) and basic reproduction number $\mathcal{R}_0(t)$ (c). \textbf{d} Epidemic peak $i_{\max}$ as a function of the initial basic reproduction number ${\cal R}_0$ for different values of $\mathcal{D}$, revealing an {abrupt epidemic transition} induced by infectivity evolution. In panels \textbf{b} and \textbf{c}, dashed lines show the theoretical predictions from Eqs. (\ref{eq:initial})-(\ref{eq:superexp}), and we set ${\cal R}_0=2$. In panels \textbf{b}, \textbf{c} and \textbf{d}, we set $i_0=10^{-3}$, $\mu=1/7$, $k=10$ and $\Delta\lambda=10^{-3}$, and stop simulations if $i(t)<10^{-4}$.}
    \label{fig:Fig1}
\end{figure}

\noindent\textit{Modeling epidemics with infectivity evolution.—}
We consider a minimal extension of the SIR model in which individuals are either susceptible (S), infected (I), or recovered (R), and where infectivity is an evolving trait (see Fig. \ref{fig:Fig1}a). For contagions, we assume that infected individuals make $k$ contacts per time step, transmitting their associated strain to a susceptible contact with a probability $\lambda$, hereafter referred to as infectivity of that strain. We assume a competitive exclusion regime, therefore neglecting coinfections by multiple strains. Following \cite{rouzine2018antigenic,zhang2022epidemic,chardes2023evolutionary,soriano2025eco}, infectivity evolves in a one-dimensional trait space via symmetric mutations of size  $\pm\Delta\lambda$ occurring at rate $D$. Finally, infected individuals recover at a rate $\mu$, entering into the R compartment.
The epidemic state of our system is characterized by the probability density of infected individuals across the traits space $\rho_I (\lambda,t)$ and the global fraction of recovered $r(t)$ individuals. Taking into account the rules described above, the time evolution of these quantities read:

\begin{align}
\frac{\partial \rho_I(\lambda,t)}{\partial t}
&=
k\,\lambda\,\rho_I(\lambda,t)\,s(t)
-\mu\,\rho_I(\lambda,t)
+\frac{\mathcal D}{2}\,
\frac{\partial^2 \rho_I(\lambda,t)}{\partial \lambda^2},\label{eq:model}
\\
\frac{dr(t)}{dt}
&=
\mu i(t),
\label{eq:model_R}
\end{align}
where $i(t)=\int d\lambda\,\rho_I(\lambda,t)$ and $s(t)=1-i(t)-r(t)$ are the disease prevalence and the fraction of susceptible individuals at time $t$ respectively. For mathematical convenience, we have used an effective diffusion rate $\mathcal D = D(\Delta\lambda)^2$, which remains invariant under rescalings of $D$ and $\Delta\lambda$ that keep $D(\Delta\lambda)^2$ constant (see Supplementary Note 1).

Eq. (\ref{eq:model}) is a reaction–diffusion equation in trait space, structurally equivalent to replicator–mutator dynamics under time-dependent selection coefficients (see Supplementary Note 1). In particular, the first term selects for strains with larger $\lambda$ through transmission, mediated by the depletion of the pool of susceptible individuals $s(t)$ throughout the dynamics. Conversely, the last term shows how mutation dynamics is translated into a diffusion of infected individuals in the trait space.
\medskip

\noindent\textit{Effect of infectivity evolution on epidemic trajectories.—} 
To characterize the epidemic trajectories, we monitor the time evolution of both the epidemic prevalence $i(t)$ and the associated basic reproduction number defined as: 
\begin{equation}
\mathcal{R}_0(t)=\frac{k\,\bar\lambda(t)}{\mu},
\label{eq:R0}
\end{equation}
with the average infectivity
\begin{equation}
    \bar\lambda(t)=\frac{\int d\lambda\,\lambda\,\rho_I(\lambda,t)}{i(t).}
    \label{eq:avg_infectivity}
\end{equation} %
Unless otherwise stated, both quantities are obtained from the numerical integration of Eqs.~(\ref{eq:model})-(\ref{eq:model_R}), assuming an initial condition where a single strain with basic reproduction number ${\cal R}_0=2$ affects a tiny fraction $i_0=10^{-3}$ of the population, with $\mu=1/7$~days$^{-1}$ and $k=10$ contacts per day. For evolution, we assume a discrete representation of the trait space with $\Delta\lambda=10^{-3}$ (see Supplementary Note 1). %
\medskip

Infectivity evolution qualitatively reshapes epidemic trajectories, inducing a super-exponential growth, advancing the epidemic peak and increasing its magnitude (Fig.~\ref{fig:Fig1}b). Moreover, Fig.~\ref{fig:Fig1}c shows that infectivity evolution presents an initial superlinear growth followed by a bending which is slowed down at later stages. Both results can be understood analytically from the dynamics of the average infectivity.
In the Supplementary Note 2, we show that differentiating Eq.~(\ref{eq:avg_infectivity}) leads to 
\begin{equation}
  \frac{d{\bar{\lambda}}}{dt}=ks(t)\,\mathrm{Var}(\lambda)\ ,
  \label{eq:derivada}
\end{equation}
where $\mathrm{Var}(\lambda)$ is the variance of $\lambda$ values considering the prevalence distribution across traits $\rho_I (\lambda,t)$. The diffusive contribution yields $\mathrm{Var}(\lambda)\simeq\mathcal{D}t$. For early-time dynamics, one can assume that $s(t)\simeq 1$. Combining both assumptions, we can integrate Eq.~(\ref{eq:derivada}), yielding:

\begin{eqnarray}
\bar\lambda(t)|_{t\rightarrow0} = \lambda_0+ \frac{1}{2}k \mathcal{D}t^2,
\label{eq:initial}
\end{eqnarray}
which directly translates into a super-exponential growth of the prevalence,
\begin{equation}
\label{eq:superexp}
i(t)\big|_{t\to 0}
=
i_0 \exp\!\left[
\mu\left({\cal R}_0-1\right)t
+\frac{k^2 \mathcal{D}}{6}\,t^3
\right].
\end{equation}
Note that infectivity evolution introduces an additional cubic term~\cite{zhang2022epidemic} in the growth exponent compared to the standard SIR model. Equations~(\ref{eq:initial})–(\ref{eq:superexp}) accurately capture the early-time dynamics in Fig.~\ref{fig:Fig1}b–c. As the epidemic progresses, the susceptible depletion slows down the growth of $\bar\lambda(t)$, producing the bending observed in Fig.~\ref{fig:Fig1}c
(see Appendix A for further details).

\medskip

\noindent\textit{Effect of infectivity evolution on epidemic control.—}
The epidemic peak $i_{\max}$ is a pivotal quantity for mathematical epidemiology, as it indicates the highest pressure on health systems during epidemic outbreaks. Fig.~\ref{fig:Fig1}d shows how the epidemic transition of the SIR model is strongly altered by evolution. While in the absence of evolution, i.e. $D=0$, epidemic outbreaks occur through a second-order transition at ${\cal R}_0=1$, increasing $D$ gives rise to abrupt epidemic transitions occurring for lower ${\cal R}_0$ values. This occurs because infectivity evolution enhances transmissibility during the outbreak, effectively pushing the dynamics into the supercritical region of the classical SIR phase diagram even when starting from marginally transmissible strains~\cite{zhang2022epidemic}. Note that the transition at lower finite ${\cal R}_0$ values appears because we assume finite populations and set a prevalence threshold $i(t)=10^{-4}$ below which the outbreak becomes extinct; in infinite populations, the active phase extends down to \({\cal R}_0=0\) for finite \(\mathcal{D}\) {(see Supplementary Note 3)}.

Moreover, Fig.~\ref{fig:Fig1}d also reveals the non-monotonic dependence of \(i_{\max}\) with ${\cal R}_0$, meaning that mildly infectious pathogens can lead to worse epidemic scenarios due to evolution. This occurs because those viruses start evolving at low prevalence values for longer times without depleting the pool of susceptible individuals, allowing the infectivity distribution to broaden and generate high-\(\lambda\) variants that are later selected by transmission {(see Supplementary Note 3)}.
\medskip

\begin{figure}
    \centering
    \includegraphics[width=1\linewidth]{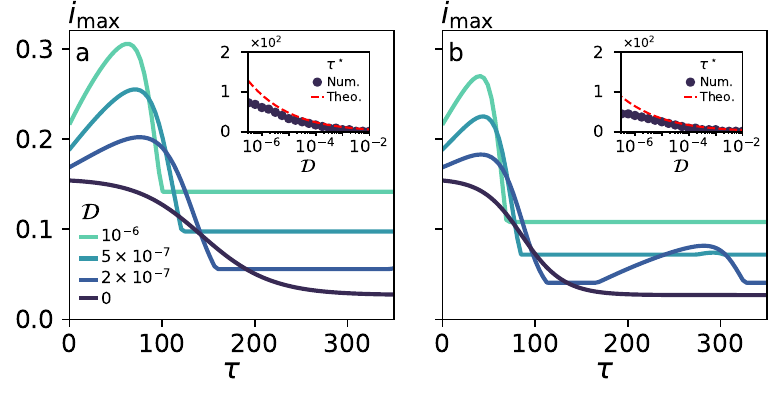}
    \caption{\textbf{Effect of infectivity evolution on epidemic control}. 
    \textbf{a} Epidemic peak $i_{\max}$ under $k-$ or $\lambda-$control as a function of the intervention time $\tau$ for different values of the effective diffusion rate $\mathcal{D}$. 
    \textbf{b} Same as in \textbf{a} for $\mu-$control.
    In both panels, inset represents the theoretical (red dashed line) prediction obtained with Eq. (\ref{eq:tau_star}) and the numerical (blue dots) values for the peak of the critical intervention time yielding the highest $i_{\max}$ value. In both panels we set $i_0=10^{-3}$, ${\cal R}_0=2$, $\mu=1/7$, $k=10$ and $\varepsilon=0.6$. %
    }
    \label{fig:Fig2}
\end{figure}

The latter result represents a cautionary tale for the short-term benefits of control policies, as lowering the basic reproduction number can lead to a worse epidemic scenario. We now tackle this issue by addressing the impact of control policies on the epidemic curves of our model. Our interventions are characterized by two parameters~\cite{morris2021optimal}: the duration of the intervention $\tau$ and its strength $\varepsilon$. Namely, our intervention rescales the reproduction number, i.e. $\mathcal{R}_0(t)\to\varepsilon \mathcal{R}_0(t)$, with $\varepsilon\in(0,1)$, over the intervention time $\tau$. Eq.~(\ref{eq:R0}) shows that such goal can be achieved through three different strategies: {\em (i)} $\lambda-$control (reducing transmissibility as $\lambda\to\varepsilon\lambda$), {\em (ii)} $k-$control (reducing contacts as $k\to\varepsilon k$) and {\em iii)} $\mu-$control (shortening the infectious period via  ($\mu\to\mu/\varepsilon$). While the first two strategies are related to inter-host transmission events, the last intervention aims at accelerating viral clearance within hosts. Unless otherwise stated, we set $\varepsilon=0.6$ and treat $k-$control and $\lambda-$control jointly, which are equivalent within the model (see Supplementary Note 4).

In the absence of infectivity evolution, all policies monotonically reduce the epidemic peak as the intervention duration $\tau$ increases (see Fig.~\ref{fig:Fig2}a for $k-$ or $\lambda-$control and Fig.~\ref{fig:Fig2}b for $\mu-$control), implying that longer interventions are always beneficial. When evolution is considered, however, $i_{\max}$ becomes a non-monotonic function of $\tau$, initially increasing and then decreasing after reaching a maximum. {This non-monotonic dependence for short durations of the control policies also appears for the attack rate $r_\infty$ (see Supplementary Note 5).} Consequently, lifting a control policy too early may result in a larger epidemic peak than in the absence of intervention ($\tau=0)$. %
Note that all the curves eventually converge to plateaus, indicating that the peak of contagions $i_{max}$ occurs during the controlled phase and not after lifting the intervention.

The non-monotonic dependency of \(i_{\max}\) on $\tau$ can be captured analytically. 
As derived in Supplementary Note 4, the early-time evolution of the average infectivity under intervention is
\begin{equation}
\bar{\lambda}(t,\varepsilon)|_{t\rightarrow0}=\lambda_0+\frac{1}{2}\,\alpha k \mathcal{D}\,t^2,
\label{eq:10}
\end{equation}
where $\alpha=\varepsilon$ for $k,\lambda-$control, and $\alpha=1$ for
$\mu-$control. As a consequence, the prevalence dynamics follows
\begin{equation}
i(t,\varepsilon)|_{t\to 0}=
i_0
\exp\!\left[
\frac{\mu\alpha}{\varepsilon} \left(\varepsilon {\cal R}_0-1 \right)t
+
\frac{\alpha^2 k^2}{6}\,
\mathcal D\,t^3
\right].
\label{eq:SM5_i_tau_general}
\end{equation}

We can estimate the peak observed after lifting the intervention at $t=\tau$, hereafter denoted by $i_{\max}(\tau)$, by introducing $i(\tau)$ and $\lambda (\tau)$ in the usual expression of the epidemic peak of the SIR model (see Appendix~B and Supplementary Note~4 for details). Differentiating $i_{\max}$ with respect to $\tau$ yields:
\begin{equation}
\label{eq:derivative_evo}
\frac{d i_{\max}}{d\tau}
=
-\mu\,i(\tau)\,\alpha\frac{(1-\varepsilon)}{\varepsilon}
+
\frac{\alpha k^2}{\mu}\,\mathcal D\,\tau\,
\frac{\ln\!\big(\mathcal {\cal R}_0(\tau)s(\tau)\big)}{\mathcal {\cal R}_0(\tau)^2}.
\end{equation}
This expression makes explicit the competition between epidemiological suppression (first term) and evolutionary amplification (second term) for short intervention durations. In particular, given an intervention duration $\tau$, the sign of the derivative changes at a critical diffusion strength ${\cal D}_c$ (detailed explicitly in the Appendix C and in Supplementary Note 4), what pinpoints that interventions improve (worsen) the epidemic scenario for viruses with $\mathcal{D}<\mathcal{D}_c$ ( $\mathcal{D}>\mathcal{D}_c$).

Considering ${\cal D}>{\cal D}_c$, we can also estimate the duration of the intervention $\tau^\star$ producing the worst epidemic scenario, i.e. the maximum value of the epidemic peak $i_{\max}$. Imposing  $\dfrac{d i_{\max}}{d\tau}\Bigr|_{\tau=\tau^*}=0$ yields (see Supplementary Note 4 for details):

\begin{equation}
\label{eq:tau_star}
\tau^\star \simeq 
\left[
-\frac{1}{3B\mathcal D}\,
W_{-1}\!\left(
-\frac{3B\mu^3(1-\varepsilon)^3 i_0^3}{\varepsilon^3C^3\,\mathcal D^2}
\right)
\right]^{1/3},
\end{equation}
with $B=\alpha^2k^2/6$ and $C=k^2\ln {\cal R}_0/(\mu {\cal R}_0^2)$. $W_{-1}$ denotes one of the branches of the Lambert function $W$, which fulfills $W(z) e^{W(z)}=z$. %
The resulting analytical predictions are in quantitative agreement with numerical simulations (see insets in Fig.~2), particularly for large values of \(\mathcal D\), where the characteristic dynamical timescale is reduced due to the evolution of the virus under control and, therefore, evolution after lifting the intervention can be neglected. The latter does not occur for lower \(\mathcal D\) values, where post-lifting evolutionary changes crucially determine $\tau^*$ values.
For completeness, in the Supplementary Note 6 we assess the influence of the basic reproduction number ${\cal R}_0$, the policy strength $\varepsilon$, the intervention activation threshold (both triggered by time delay or prevalence), and a possible negative bias towards lower infectivity values, showing that the nonmonotonic behavior is robust.
\medskip

\noindent\textit{Asymmetric effect of control policies shaping inter-host or within-host dynamics.—}
Comparing Figs.~\ref{fig:Fig2}a-b, we observe a difference between $k,\lambda-$control, and $\mu-$control. In the absence of infectivity evolution, {\em i.e.}, the usual SIR model, shortening the infectious period yields a smaller post-lifting peak than reducing transmissibility. We observe that infectivity evolution amplifies this effect for short interventions whereas it might revert it for longer interventions, as highlighted by the secondary peak appearing for ${\cal D}=2\times 10^{-7}$ in Fig.~\ref{fig:Fig2}b.

\begin{figure}[t!]
    \centering
    \includegraphics[width=1\linewidth]{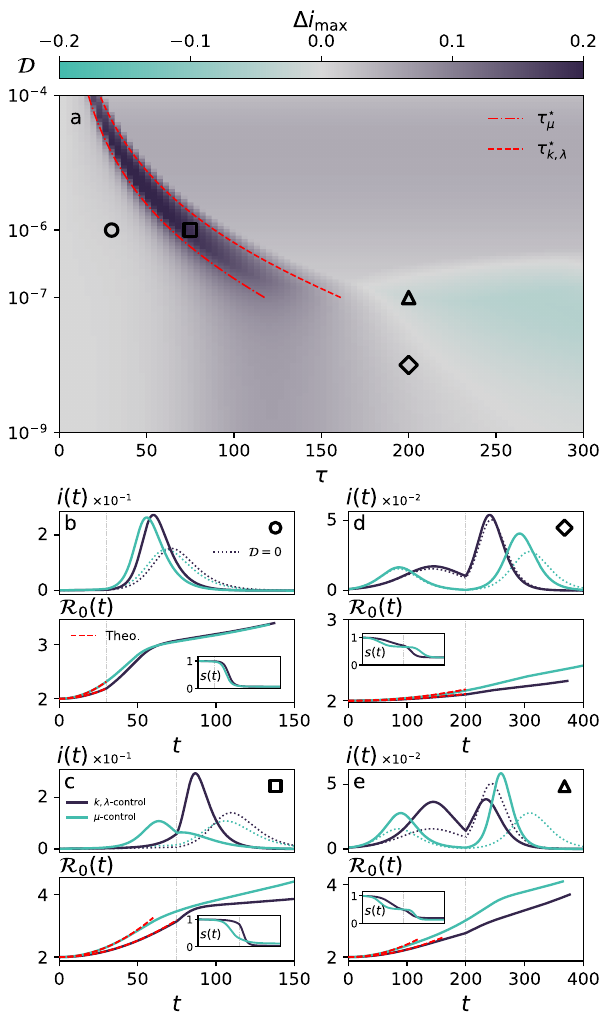}
    \caption{\textbf{Asymmetric effect of control policies.}
    \textbf{a} Difference in epidemic peak $\Delta i_{\max}$ between the $k,\lambda-$control and $\mu-$control as a function of $\tau$ and $\mathcal{D}$. Dashed curves indicate the theoretical predictions for the critical intervention durations $\tau^{\star}$ associated with each strategy.
    \textbf{b}-\textbf{e} Temporal evolution of the epidemic prevalence $i(t)$ (top row) and the corresponding time-varying basic reproduction number $\mathcal{R}_0(t)$ (bottom row) for (b) $\tau=30$ and $\mathcal{D}=10^{-6}$, (c) $\tau=75$ and $\mathcal{D}=10^{-6}$, (d) $\tau=200$ and $\mathcal{D}=10^{-8}$ and (e) $\tau=200$ and $\mathcal{D}=10^{-7}$.
    Panels \textbf{b}-\textbf{e} include the theoretical prediction for the early-time dynamics (red dashed line) and, in the inset, the evolution of the susceptible population $s(t)$. Moreover, we stop simulations if $i(t)<10^{-4}$.
    In all panels, we set $i_0=10^{-3}$, ${\cal R}_0=2$, $\mu=1/7$, $k=10$ and $\varepsilon=0.6$. Colors distinguish control strategies, as indicated in panel \textbf{c}, and vertical lines indicate lifting times. Symbols in panel \textbf{a} correspond to the parameter choices shown in panels \textbf{b}-\textbf{e}.}
    \label{fig:Fig3}
\end{figure}

The impact of the difference between interventions is made explicit in Fig.~\ref{fig:Fig3}a, which shows $\Delta i_{\max}=i_{\max}^{(k,\lambda)}-i_{\max}^{(\mu)}$. Considering a constant ${\cal D}$ value, e.g. ${\cal D}=10^{-6}$, we first observe that increasing $\tau$ yields a transition from a similar outcome of both interventions ($\Delta i_{\max}\simeq0$) to a substantially larger epidemic peak after $k,\lambda-$control compared to the one observed after $\mu-$control. Note that this transition also appears in absence of evolution, i.e. very low {$\mathcal D$} values, but evolution exacerbates it and anticipates its onset.

To gain insights into this transition, we represent the epidemic trajectories $i(t)$ along with the evolution of the basic reproduction number ${\cal R}_0 (t)$ and the pool of susceptible individuals $s(t)$ for $\tau=30$ days (Fig.~\ref{fig:Fig3}b) and $\tau=75$ days (Fig.~\ref{fig:Fig3}c). Comparing the curves of susceptible individuals, we can observe how the acceleration of the epidemic time scales triggered by $\mu-$control leads to a faster depletion of the pool of susceptible individuals. While this faster depletion is not substantial for very short interventions~(Fig.~\ref{fig:Fig3}b), it becomes evident once the peak of the $\mu-$control epidemic curves has been reached~(Fig.~\ref{fig:Fig3}c). In that case, the exhausted pool of susceptible individuals prevents the emergence of another major outbreak, as observed in the case of the $k,\lambda$-control. Using this intuition, the region of strong asymmetry should be delimited by those intervention times yielding the highest epidemic peaks for both interventions, i.e. $\tau_\mu^\star$ and $\tau_{k,\lambda}^\star$. Thus, we can compute them with Eq.~(\ref{eq:tau_star}), using the $\alpha$ value of each policy. The red dashed lines in Fig.~\ref{fig:Fig3}a show a fair agreement between our theoretical bounds for this region and the results from numerical simulations.

Fig.~\ref{fig:Fig3}a also illustrates that longer interventions and moderate evolutionary rates revert the asymmetric outcome of control policies, producing a higher peak for the $\mu$-control strategy, i.e. $\Delta i_{\max} \leq 0$. To understand this phenomenon, we consider $\tau=200$ and represent the epidemic and evolutionary curves assuming ${\cal D}=10^{-8}$ (Fig.~\ref{fig:Fig3}d) and ${\mathcal D}=10^{-7}$ (Fig.~\ref{fig:Fig3}e). In both cases, $s(t)$ does not allow us to explain the unequal outcome, as the pool of susceptible individuals is only partially depleted and of similar size for each strategy. We should instead focus on how the different control policies shape the evolution of the virus. While $\mu-$control yields the same early-time evolution of ${\mathcal R}_0$ as in the uncontrolled case, $k,\lambda-$control slows down viral evolution. Eq.~(\ref{eq:10}) captures this difference, encoded in the parameter $\alpha$, with a fair agreement between the theoretical lines and the numerical results at early stages of the outbreak.

Therefore, for intermediate evolutionary speeds, i.e. $\mathcal D\sim 10^{-7}$, the pool of susceptible agents is large and of similar size across strategies at $t=\tau$. However, viruses under $\mu-$control have much higher infectiousness than those under $k,\lambda-$control, thus explaining the higher epidemic peak in the secondary outbreak~\cite{castioni2024rebound}. Note that this behavior does not appear for faster evolutionary rates as they induce a much greater depletion of the susceptible population, not observing any secondary peak.

\medskip

\noindent\textit{Conclusions.—} 
We have shown that allowing pathogen infectivity to evolve reshapes in a fundamental way both epidemic dynamics and the effectiveness of control strategies. The interplay between contagion (selection) and mutation (diffusion) induces superexponential early-time prevalence growth. {This dynamical behavior has been theoretically predicted for epidemics \cite{zhang2022epidemic} and reported in empirical data for influenza outbreaks~\cite{scarpino2016effect}, whereby it was attributed to a different underlying mechanism. Accelerated growth rates have also been observed in other biological dynamics~\cite{durrett2010evolutionary} such as tumor growth~\cite{ocana2023growth} or bacterial proliferation~\cite{cylke2023super}.} {The model also displays abrupt epidemic transitions~\cite{zhang2022epidemic},} positioning trait evolution as an additional pathway towards explosive transitions in contagion dynamics \cite{d2019explosive,lamata2024pathways,lamata2023collapse}.

When control measures are incorporated, these evolutionary effects lead to qualitatively different and counterintuitive outcomes. In particular, both the attack rate and the epidemic peak depend nonmonotonically on intervention duration, implying that prematurely lifted interventions may worsen epidemic outcomes relative to doing nothing.
{This non-monotonic behavior is unimodal for the epidemic peak but presents multiple local maxima for the attack rate (see Supplementary Note 5); therefore, multiple epidemiological indicators should be computed for a complete understanding of how control policies shape the impact and evolution of epidemic outbreaks~\cite{rozan2026analysis}.}

Furthermore, we identified an asymmetry between control strategies: interventions acting on transmission parameters and those shortening the infectious period are no longer equivalent in the presence of infectivity evolution. Shortening the infectious period, e.g. by administering drugs, accelerates epidemic timescales without slowing infectivity evolution, whereas acting on inter-host transmission slows both prevalence and infectivity growth.
As a result, the relative effectiveness of control strategies may be reversed: policies that appear optimal in non-evolving settings can become counterproductive once evolutionary effects are taken into account, particularly when interventions are lifted after the first epidemic peak.

Overall, {using this minimal model, we have revealed a natural multiscale coupling between mutation dynamics within hosts and epidemic control at the population level, which determines the selection of newly emerging variants and, consequently, viral evolution. Despite the simplicity of the model,} our findings are quite general and expected to appear in more biologically-grounded models including complementary evolutionary pathways such as antigenic drift~\cite{sasaki2022antigenic,soriano2025eco,lamata2025genotype,williams2021localization} or time-evolving recovery rates, exhaustive descriptions of within-host dynamics~\cite{mideo2008linking,fabre2012modelling} or evolutionary trade-offs constraining viral evolution~\cite{alizon2008transmission,de2008virulence,acevedo2019virulence}.

\medskip

\noindent\textit{Acknowledgments—}
S.L.O and J.G.G. acknowledge financial support from the Departamento de Industria e Innovaci\'on del Gobierno de Arag\'on y Fondo Social Europeo (FENOL group grant E36-23R) and from Ministerio de Ciencia e Innovaci\'on (grant PID2023-147734NB-I00). S.L.O. acknowledges financial support from Gobierno de Aragón through a doctoral fellowship. A.A. and D.S.-P acknowledge Spanish Ministerio de Ciencia e Innovaci\'on (PID2024-158120NB-C21). AA acknowledges Generalitat de Catalunya (2021SGR-00633), Universitat Rovira i Virgili (2023PFR-URV-00633), the European Union’s Horizon Europe Programme under the CREXDATA project (grant agreement no.\ 101092749), ICREA Academia, the James S.\ McDonnell Foundation (Grant N.\ 220020325), and the Joint Appointment Program at Pacific Northwest National Laboratory (PNNL). PNNL is a multi-program national laboratory operated for the U.S.\ Department of Energy (DOE) by Battelle Memorial Institute under Contract No.\ DE-AC05-76RL01830. 
\medskip

\noindent\textit{Code availability—}
The code is available at https://github.com/santiagolaot/SIRevolution.
\medskip

\bibliography{biblio}

@article{rouzine2018antigenic,
  title={Antigenic evolution of viruses in host populations},
  author={Rouzine, Igor M and Rozhnova, Ganna},
  journal={PLoS pathogens},
  volume={14},
  number={9},
  pages={e1007291},
  year={2018},
  publisher={Public Library of Science San Francisco, CA USA}
}

@article{zhang2022epidemic,
  title={Epidemic spreading under mutually independent intra-and inter-host pathogen evolution},
  author={Zhang, Xiyun and Ruan, Zhongyuan and Zheng, Muhua and Zhou, Jie and Boccaletti, Stefano and Barzel, Baruch},
  journal={Nature communications},
  volume={13},
  number={1},
  pages={6218},
  year={2022},
  publisher={Nature Publishing Group UK London}
}

@article{chardes2023evolutionary,
  title={Evolutionary stability of antigenically escaping viruses},
  author={Chard{\`e}s, Victor and Mazzolini, Andrea and Mora, Thierry and Walczak, Aleksandra M},
  journal={Proceedings of the National Academy of Sciences},
  volume={120},
  number={44},
  pages={e2307712120},
  year={2023},
  publisher={National Academy of Sciences}
}

@article{morris2021optimal,
  title={Optimal, near-optimal, and robust epidemic control},
  author={Morris, Dylan H and Rossine, Fernando W and Plotkin, Joshua B and Levin, Simon A},
  journal={Communications Physics},
  volume={4},
  number={1},
  pages={78},
  year={2021},
  publisher={Nature Publishing Group UK London}
}

@article{lamata2024pathways,
  title={Pathways to discontinuous transitions in interacting contagion dynamics},
  author={Lamata-Ot{\'\i}n, Santiago and G{\'o}mez-Garde{\~n}es, Jes{\'u}s and Soriano-Pa{\~n}os, David},
  journal={Journal of Physics: Complexity},
  volume={5},
  number={1},
  pages={015015},
  year={2024},
  publisher={IOP Publishing}
}

@article{lamata2023collapse,
  title={Collapse transition in epidemic spreading subject to detection with limited resources},
  author={Lamata-Ot{\'\i}n, Santiago and Reyna-Lara, Adriana and Soriano-Pa{\~n}os, David and Latora, Vito and G{\'o}mez-Garde{\~n}es, Jes{\'u}s},
  journal={Physical Review E},
  volume={108},
  number={2},
  pages={024305},
  year={2023},
  publisher={APS}
}

@article{sasaki2022antigenic,
  title={Antigenic escape selects for the evolution of higher pathogen transmission and virulence},
  author={Sasaki, Akira and Lion, S{\'e}bastien and Boots, Mike},
  journal={Nature ecology \& evolution},
  volume={6},
  number={1},
  pages={51--62},
  year={2022},
  publisher={Nature Publishing Group UK London}
}

@article{lamata2025genotype,
  title={Genotype networks drive oscillating endemicity and epidemic trajectories in viral evolution},
  author={Lamata-Ot{\'\i}n, Santiago and Rotita-Ion, Octavian C and Arenas, Alex and Soriano-Pa{\~n}os, David and G{\'o}mez-Garde{\~n}es, Jes{\'u}s},
  journal={Communications Physics},
  year={2025},
  publisher={Nature Publishing Group UK London}
}

@article{pastor2015epidemic,
  title={Epidemic processes in complex networks},
  author={Pastor-Satorras, Romualdo and Castellano, Claudio and Van Mieghem, Piet and Vespignani, Alessandro},
  journal={Reviews of modern physics},
  volume={87},
  number={3},
  pages={925--979},
  year={2015},
  publisher={APS}
}

@book{keeling2008modeling,
  title={Modeling infectious diseases in humans and animals},
  author={Keeling, Matt J and Rohani, Pejman},
  year={2008},
  publisher={Princeton university press}
}

@article{kermack1927contribution,
  title={A contribution to the mathematical theory of epidemics},
  author={Kermack, William Ogilvy and McKendrick, Anderson G},
  journal={Proceedings of the royal society of london. Series A, Containing papers of a mathematical and physical character},
  volume={115},
  number={772},
  pages={700--721},
  year={1927},
  publisher={The Royal Society London}
}

@article{hethcote2000mathematics,
  title={The mathematics of infectious diseases},
  author={Hethcote, Herbert W},
  journal={SIAM review},
  volume={42},
  number={4},
  pages={599--653},
  year={2000},
  publisher={SIAM}
}

@article{flaxman2020estimating,
  title={Estimating the effects of non-pharmaceutical interventions on COVID-19 in Europe},
  author={Flaxman, Seth and Mishra, Swapnil and Gandy, Axel and Unwin, H Juliette T and Mellan, Thomas A and Coupland, Helen and Whittaker, Charles and Zhu, Harrison and Berah, Tresnia and Eaton, Jeffrey W and others},
  journal={Nature},
  volume={584},
  number={7820},
  pages={257--261},
  year={2020},
  publisher={Nature Publishing Group UK London}
}

@article{hsiang2020effect,
  title={The effect of large-scale anti-contagion policies on the COVID-19 pandemic},
  author={Hsiang, Solomon and Allen, Daniel and Annan-Phan, S{\'e}bastien and Bell, Kendon and Bolliger, Ian and Chong, Trinetta and Druckenmiller, Hannah and Huang, Luna Yue and Hultgren, Andrew and Krasovich, Emma and others},
  journal={Nature},
  volume={584},
  number={7820},
  pages={262--267},
  year={2020},
  publisher={Nature Publishing Group UK London}
}

@article{kissler2020projecting,
  title={Projecting the transmission dynamics of SARS-CoV-2 through the postpandemic period},
  author={Kissler, Stephen M and Tedijanto, Christine and Goldstein, Edward and Grad, Yonatan H and Lipsitch, Marc},
  journal={Science},
  volume={368},
  number={6493},
  pages={860--868},
  year={2020},
  publisher={American Association for the Advancement of Science}
}

@article{arenas2020modeling,
  title={Modeling the spatiotemporal epidemic spreading of COVID-19 and the impact of mobility and social distancing interventions},
  author={Arenas, Alex and Cota, Wesley and G{\'o}mez-Garde{\~n}es, Jes{\'u}s and G{\'o}mez, Sergio and Granell, Clara and Matamalas, Joan T and Soriano-Pa{\~n}os, David and Steinegger, Benjamin},
  journal={Physical Review X},
  volume={10},
  number={4},
  pages={041055},
  year={2020},
  publisher={APS}
}

@article{di2021optimal,
  title={Optimal timing of one-shot interventions for epidemic control},
  author={Di Lauro, Francesco and Kiss, Istv{\'a}n Z and Miller, Joel C},
  journal={PLOS Computational Biology},
  volume={17},
  number={3},
  pages={e1008763},
  year={2021},
  publisher={Public Library of Science San Francisco, CA USA}
}

@article{bliman2021optimal,
  title={Optimal immunity control and final size minimization by social distancing for the SIR epidemic model},
  author={Bliman, Pierre-Alexandre and Duprez, Michel and Privat, Yannick and Vauchelet, Nicolas},
  journal={Journal of Optimization Theory and Applications},
  volume={189},
  number={2},
  pages={408--436},
  year={2021},
  publisher={Springer}
}

@article{ketcheson2021optimal,
  title={Optimal control of an SIR epidemic through finite-time non-pharmaceutical intervention},
  author={Ketcheson, David I},
  journal={Journal of mathematical biology},
  volume={83},
  number={1},
  pages={7},
  year={2021},
  publisher={Springer}
}

@article{perkins2020optimal,
  title={Optimal control of the COVID-19 pandemic with non-pharmaceutical interventions},
  author={Perkins, T Alex and Espa{\~n}a, Guido},
  journal={Bulletin of mathematical biology},
  volume={82},
  number={9},
  pages={118},
  year={2020},
  publisher={Springer}
}

@article{grenfell2004unifying,
  title={Unifying the epidemiological and evolutionary dynamics of pathogens},
  author={Grenfell, Bryan T and Pybus, Oliver G and Gog, Julia R and Wood, James LN and Daly, Janet M and Mumford, Jenny A and Holmes, Edward C},
  journal={science},
  volume={303},
  number={5656},
  pages={327--332},
  year={2004},
  publisher={American Association for the Advancement of Science}
}

@book{anderson1991infectious,
  title={Infectious diseases of humans: dynamics and control},
  author={Anderson, Roy M and May, Robert M},
  year={1991},
  publisher={Oxford university press}
}

@article{day2007applying,
  title={Applying population-genetic models in theoretical evolutionary epidemiology},
  author={Day, Troy and Gandon, Sylvain},
  journal={Ecology Letters},
  volume={10},
  number={10},
  pages={876--888},
  year={2007},
  publisher={Wiley Online Library}
}

@article{lion2018beyond,
  title={Beyond R0 maximisation: on pathogen evolution and environmental dimensions},
  author={Lion, S{\'e}bastien and Metz, Johan AJ},
  journal={Trends in ecology \& evolution},
  volume={33},
  number={6},
  pages={458--473},
  year={2018},
  publisher={Elsevier}
}

@article{williams2021localization,
  title={Localization, epidemic transitions, and unpredictability of multistrain epidemics with an underlying genotype network},
  author={Williams, Blake JM and St-Onge, Guillaume and H{\'e}bert-Dufresne, Laurent},
  journal={PLoS Computational Biology},
  volume={17},
  number={2},
  pages={e1008606},
  year={2021},
  publisher={Public Library of Science San Francisco, CA USA}
}

@article{soriano2025eco,
  title={Eco-evolutionary constraints for the endemicity of rapidly evolving viruses},
  author={Soriano-Pa{\~n}os, David},
  journal={PRX Life},
  volume={3},
  pages={043001},
  year={2025},
  publisher={APS}
}

@article{dellicour2018phylodynamic,
  title={Phylodynamic assessment of intervention strategies for the West African Ebola virus outbreak},
  author={Dellicour, Simon and Baele, Guy and Dudas, Gytis and Faria, Nuno R and Pybus, Oliver G and Suchard, Marc A and Rambaut, Andrew and Lemey, Philippe},
  journal={Nature communications},
  volume={9},
  number={1},
  pages={2222},
  year={2018},
  publisher={Nature Publishing Group UK London}
}

@article{van2021controlling,
  title={Controlling long-term SARS-CoV-2 infections can slow viral evolution and reduce the risk of treatment failure},
  author={Van Egeren, Debra and Novokhodko, Alexander and Stoddard, Madison and Tran, Uyen and Zetter, Bruce and Rogers, Michael S and Joseph-McCarthy, Diane and Chakravarty, Arijit},
  journal={Scientific reports},
  volume={11},
  number={1},
  pages={22630},
  year={2021},
  publisher={Nature Publishing Group UK London}
}

@article{iwasa2003evolutionary,
  title={Evolutionary dynamics of escape from biomedical intervention},
  author={Iwasa, Yoh and Michor, Franziska and Nowak, Martin A},
  journal={Proceedings of the Royal Society of London. Series B: Biological Sciences},
  volume={270},
  number={1533},
  pages={2573--2578},
  year={2003},
  publisher={The Royal Society}
}

@article{hartfield2014epidemiological,
  title={Epidemiological feedbacks affect evolutionary emergence of pathogens},
  author={Hartfield, Matthew and Alizon, Samuel},
  journal={The American Naturalist},
  volume={183},
  number={4},
  pages={E105--E117},
  year={2014},
  publisher={University of Chicago Press Chicago, IL}
}

@article{iwasa2004evolutionary,
  title={Evolutionary dynamics of invasion and escape},
  author={Iwasa, Yoh and Michor, Franziska and Nowak, Martin A},
  journal={Journal of Theoretical Biology},
  volume={226},
  number={2},
  pages={205--214},
  year={2004},
  publisher={Elsevier}
}

@article{hartfield2015within,
  title={Within-host stochastic emergence dynamics of immune-escape mutants},
  author={Hartfield, Matthew and Alizon, Samuel},
  journal={PLoS computational biology},
  volume={11},
  number={3},
  pages={e1004149},
  year={2015},
  publisher={Public Library of Science San Francisco, CA USA}
}

@article{meehan2020probability,
  title={On the probability of strain invasion in endemic settings: accounting for individual heterogeneity and control in multi-strain dynamics},
  author={Meehan, Michael T and Cope, Robert C and McBryde, Emma S},
  journal={Journal of theoretical biology},
  volume={487},
  pages={110109},
  year={2020},
  publisher={Elsevier}
}

@article{ashby2023non,
  title={Non-pharmaceutical interventions and the emergence of pathogen variants},
  author={Ashby, Ben and Smith, Cameron A and Thompson, Robin N},
  journal={Evolution, Medicine, and Public Health},
  volume={11},
  number={1},
  pages={80--89},
  year={2023},
  publisher={Oxford University Press US}
}

@article{chen2024covid,
  title={COVID-19 pandemic interventions reshaped the global dispersal of seasonal influenza viruses},
  author={Chen, Zhiyuan and Tsui, Joseph L-H and Gutierrez, Bernardo and Busch Moreno, Simon and Du Plessis, Louis and Deng, Xiaowei and Cai, Jun and Bajaj, Sumali and Suchard, Marc A and Pybus, Oliver G and others},
  journal={Science},
  volume={386},
  number={6722},
  pages={eadq3003},
  year={2024},
  publisher={American Association for the Advancement of Science}
}

@article{chen2025disruption,
  title={Disruption of seasonal influenza circulation and evolution during the 2009 H1N1 and COVID-19 pandemics in Southeastern Asia},
  author={Chen, Zhiyuan and Tsui, Joseph L-H and Cai, Jun and Su, Shuo and Viboud, C{\'e}cile and Du Plessis, Louis and Lemey, Philippe and Kraemer, Moritz UG and Yu, Hongjie},
  journal={Nature Communications},
  volume={16},
  number={1},
  pages={475},
  year={2025},
  publisher={Nature Publishing Group UK London}
}

@article{tsimring1996rna,
  title={RNA virus evolution via a fitness-space model},
  author={Tsimring, Lev S and Levine, Herbert and Kessler, David A},
  journal={Physical review letters},
  volume={76},
  number={23},
  pages={4440},
  year={1996},
  publisher={APS}
}

@article{hindes2016epidemic,
  title={Epidemic extinction and control in heterogeneous networks},
  author={Hindes, Jason and Schwartz, Ira B},
  journal={Physical review letters},
  volume={117},
  number={2},
  pages={028302},
  year={2016},
  publisher={APS}
}

@article{fabre2012modelling,
  title={Modelling the evolutionary dynamics of viruses within their hosts: a case study using high-throughput sequencing},
  author={Fabre, Fr{\'e}d{\'e}ric and Montarry, Josselin and Coville, J{\'e}r{\^o}me and Senoussi, Rachid and Simon, Vincent and Moury, Beno{\^\i}t},
  journal={PLoS Pathogens},
  volume={8},
  number={4},
  pages={e1002654},
  year={2012},
  publisher={Public Library of Science San Francisco, USA}
}

@article{mideo2008linking,
  title={Linking within-and between-host dynamics in the evolutionary epidemiology of infectious diseases},
  author={Mideo, Nicole and Alizon, Samuel and Day, Troy},
  journal={Trends in ecology \& evolution},
  volume={23},
  number={9},
  pages={511--517},
  year={2008},
  publisher={Elsevier}
}

@article{alizon2008transmission,
  title={Transmission-recovery trade-offs to study parasite evolution},
  author={Alizon, Samuel},
  journal={The American Naturalist},
  volume={172},
  number={3},
  pages={E113--E121},
  year={2008},
  publisher={The University of Chicago Press}
}

@article{de2008virulence,
  title={Virulence-transmission trade-offs and population divergence in virulence in a naturally occurring butterfly parasite},
  author={De Roode, Jacobus C and Yates, Andrew J and Altizer, Sonia},
  journal={Proceedings of the national academy of sciences},
  volume={105},
  number={21},
  pages={7489--7494},
  year={2008},
  publisher={National Academy of Sciences}
}

@article{acevedo2019virulence,
  title={Virulence-driven trade-offs in disease transmission: A meta-analysis},
  author={Acevedo, Miguel A and Dillemuth, Forrest P and Flick, Andrew J and Faldyn, Matthew J and Elderd, Bret D},
  journal={Evolution},
  volume={73},
  number={4},
  pages={636--647},
  year={2019},
  publisher={Blackwell Publishing Inc Malden, USA}
}

@article{castioni2024rebound,
  title={Rebound in epidemic control: How misaligned vaccination timing amplifies infection peaks},
  author={Castioni, Piergiorgio and G{\'o}mez, Sergio and Granell, Clara and Arenas, Alex},
  journal={npj Complexity},
  volume={1},
  number={1},
  pages={20},
  year={2024},
  publisher={Nature Publishing Group UK London}
}

@article{d2019explosive,
  title={Explosive phenomena in complex networks},
  author={D'Souza, Raissa M and G{\'o}mez-Gardenes, Jesus and Nagler, Jan and Arenas, Alex},
  journal={Advances in Physics},
  volume={68},
  number={3},
  pages={123--223},
  year={2019},
  publisher={Taylor \& Francis}
}

@article{rozan2026analysis,
  title={Analysis of non pharmaceutical interventions with SIR epidemic models: decreasing the infection peak vs. minimizing the epidemic size},
  author={Roz{\'a}n, Eric and Kuperman, Marcelo N and Bouzat, Sebasti{\'a}n},
  journal={arXiv preprint arXiv:2604.08420},
  year={2026}
}

@article{durrett2010evolutionary,
  title={Evolutionary dynamics of tumor progression with random fitness values},
  author={Durrett, Rick and Foo, Jasmine and Leder, Kevin and Mayberry, John and Michor, Franziska},
  journal={Theoretical population biology},
  volume={78},
  number={1},
  pages={54--66},
  year={2010},
  publisher={Elsevier}
}

@article{ocana2023growth,
  title={Growth exponents reflect evolutionary processes and treatment response in brain metastases},
  author={Oca{\~n}a-Tienda, Beatriz and P{\'e}rez-Beteta, Juli{\'a}n and Jim{\'e}nez-S{\'a}nchez, Juan and Molina-Garc{\'\i}a, David and Ortiz de Mendivil, Ana and Asenjo, Beatriz and Albillo, David and P{\'e}rez-Romasanta, Luis A and Valiente, Manuel and Zhu, Luc{\'\i}a and others},
  journal={npj Systems Biology and Applications},
  volume={9},
  number={1},
  pages={35},
  year={2023},
  publisher={Nature Publishing Group UK London}
}

@article{cylke2023super,
  title={Super-exponential growth and stochastic size dynamics in rod-like bacteria},
  author={Cylke, Arianna and Banerjee, Shiladitya},
  journal={Biophysical Journal},
  volume={122},
  number={7},
  pages={1254--1267},
  year={2023},
  publisher={Elsevier}
}

@article{scarpino2016effect,
  title={The effect of a prudent adaptive behaviour on disease transmission},
  author={Scarpino, Samuel V and Allard, Antoine and H{\'e}bert-Dufresne, Laurent},
  journal={Nature Physics},
  volume={12},
  number={11},
  pages={1042--1046},
  year={2016},
  publisher={Nature Publishing Group UK London}
}

\onecolumngrid
\par\bigskip
\begin{center}
{\large\bfseries End Matter}
\end{center}
\twocolumngrid

\renewcommand{\theequation}{A\arabic{equation}}
\setcounter{equation}{0}

\noindent\textit{Appendix A: Sigmoid approximation for susceptible depletion—}
To analyze the slowdown in the growth of the average infectivity induced by susceptible depletion, we approximate the susceptible fraction by a sigmoid function centered at the epidemic peak,
$s(t)\simeq s(\infty)+\bigl(1-s(\infty)\bigr)/(1+e^{\sigma(t-t_{\mathrm{p}})})$.
Substituting this expression into Eq. (\ref{eq:derivada}) and integrating
yields an explicit approximation for the evolution of the average infectivity at intermediate times (see Supplementary Note~7). Retaining the dominant contribution for times around and beyond the peak time $t_{\mathrm{p}}$, we obtain
\begin{eqnarray}
&\bar{\lambda}(t)&\;\sim
k\mathcal{D}\\&&\times\left[
s(\infty)\frac{t^2}{2}
-(1-s(\infty))
\left(
\frac{\sigma t+1}{\sigma^2}
e^{-\sigma(t-t_{\mathrm{p}})}
\right)
\right]\nonumber,
\label{eq:A3}
\end{eqnarray}
which explicitly displays a negative transient contribution that temporarily hinders the growth of infectivity (see Fig.~\ref{fig:Fig1}c). As time increases, the exponential term vanishes and the dynamics smoothly crosses over to a quadratic growth controlled by the remaining susceptible fraction $s(\infty)$.
\medskip

\noindent\textit{Appendix B: Hybrid approximation to derive the epidemic peak—}
To assess the impact of infectivity evolution on the effectiveness of control
policies, we adopt a hybrid approximation in which the epidemic dynamics during
the intervention window $t\in[0,\tau]$ follows the early-time controlled evolution
of $i(t,\varepsilon)$ and $\bar{\lambda}(t,\varepsilon)$, while after lifting the intervention the outbreak
evolves as an uncontrolled SIR process, with initial conditions given by the state
at $t=\tau$ (see Supplementary Note~4 for details).

Under this approximation, and assuming that the epidemic peak occurs after lifting
the intervention, the post-intervention peak can be approximated using the standard
SIR result~\cite{lamata2023collapse}
\begin{equation}
i_{\max}(\tau)
\approx
1-r(\tau)
-\frac{1}{\mathcal{R}_0(\tau)}
\left[
1+\ln\!\big(\mathcal{R}_0(\tau)s(\tau)\big)
\right],
\label{eq:C1}
\end{equation}
where $\mathcal{R}_0(\tau)=k\bar{\lambda}(\tau)/\mu$ and
$s(\tau)=1-i(\tau)-r(\tau)$. Note that this expression is a strong assumption as it neglects the impact of evolution on the epidemic peak; evolution just affects the initial conditions when lifting the intervention.

Differentiating Eq.~(\ref{eq:C1}) with respect to $\tau$ and using the expressions
for the controlled dynamics at the lifting time
[Eqs.~(\ref{eq:10})–(\ref{eq:SM5_i_tau_general})] yields the general identity reported in
Eq.~(\ref{eq:derivative_evo}), which explicitly separates the direct
epidemiological contribution from the evolutionary one.
\medskip

\noindent\textit{Appendix C: Critical effective diffusion strength—}
To quantify when infectivity evolution reverses the monotonic dependence of the
epidemic peak on the intervention duration, we analyze the condition
$d i_{\max}/d\tau=0$, reported in Eq.~(\ref{eq:derivative_evo}) and derived in
Appendix~B. Setting $d i_{\max}/d\tau=0$ defines an implicit relation for the critical
effective diffusion strength,
\begin{equation}
\label{eq:Dcrit_general}
\mathcal{D}_c(\tau)
=
\frac{\mu^2(1-\varepsilon)\,i(\tau)\,\mathcal{R}_0(\tau)^2}
{\varepsilon k^2\,\tau\,\ln\!\big(\mathcal{R}_0(\tau)s(\tau)\big)}.
\end{equation}

Assuming that the epidemic prevalence remains small at the lifting time,
$i(\tau)\ll1$, we approximate $s(\tau)\simeq1$ and
$\ln\!\big(\mathcal{R}_0(\tau)s(\tau)\big)\simeq\ln {\cal R}_0$, and evaluate the logarithmic
and prefactor terms at their zeroth-order values,
$\mathcal{R}_0(\tau)\simeq {\cal R}_0$, while retaining only the leading dependence on
$\mathcal{D}$ through the exponential contribution in $i(\tau)$. As detailed in
Supplementary Note~4, under these approximations and using Eq. (\ref{eq:SM5_i_tau_general}), Eq.~(\ref{eq:Dcrit_general})
reduces to a transcendental equation that admits a closed-form solution in terms
of the Lambert $W$ function,
\begin{equation}
\label{eq:Dc_lambert_general}
\mathcal{D}_c(\tau)=
-\frac{1}{P(\tau)}\,W_{0}\!\left[-P(\tau)\,Q(\tau)\right],
\end{equation}
where $P(\tau)=\alpha^2 k^2\,\tau^3/6$ and $Q(\tau)=
(\mu^2(1-\varepsilon)\,{\cal R}_0^2)/(\varepsilon k^2\,\tau\,\ln {\cal R}_0)\;
i_0\,
\exp\!\left[\mu\alpha\left(\varepsilon {\cal R}_0-1\right)\tau/\varepsilon\right]$. In the expression above $\alpha=\varepsilon$ for $k$- or $\lambda$-control and
$\alpha=1$ for $\mu$-control.

Equation~(\ref{eq:Dc_lambert_general}) exhibits a divergence
$\mathcal{D}_c(\tau)\to\infty$ as $\tau\to0$, reflecting the fact that the hybrid
approximation accounts for infectivity evolution only during the controlled phase
and therefore does not capture the $\tau\to0$ limit. For finite intervention
durations, however, it predicts a critical effective diffusion strength such that,
for $\mathcal{D}>\mathcal{D}_c(\tau)$ the epidemic peak increases with the
intervention duration, whereas for $\mathcal{D}<\mathcal{D}_c(\tau)$ it decreases.
\medskip

\clearpage

\newpage

\renewcommand{\figurename}{Supplementary Fig.}
\renewcommand{\tablename}{Supplementary Table}
\renewcommand{\theequation}{S.\arabic{equation}}

\setcounter{equation}{0}
\setcounter{figure}{0}

\onecolumngrid

\section*{Supplementary Note 1: Coevolutionary model}

\subsection*{Mapping to replicator--mutator dynamics}

Integrating Eq.~(1) in the main text over $\lambda$ yields
\begin{equation}
\frac{d i}{dt}=\big(k s(t)\bar\lambda(t)-\mu\big)\,i(t),
\label{eq:SM_i_dot}
\end{equation}
where $i(t)=\int d\lambda\,\rho_I(\lambda,t)$ and $\bar\lambda(t)
=(\int d\lambda\,\lambda\,\rho_I(\lambda,t))/(
\int d\lambda\,\rho_I(\lambda,t))$.
We now introduce the normalized trait distribution $x(\lambda,t)=\frac{\rho_I(\lambda,t)}{i(t)}$, such that $\int d\lambda\,x(\lambda,t)=1$. Taking the time derivative of $x(\lambda,t)$,
\begin{equation}
\frac{\partial x}{\partial t}
=
\frac{1}{i}\frac{\partial \rho_I}{\partial t}
-
\frac{\rho_I}{i^2}\frac{\partial i}{\partial t},
\end{equation}
and substituting Eq.~(1) together with Eq.~(\ref{eq:SM_i_dot}),
we obtain after rearranging terms
\begin{equation}
\frac{\partial x(\lambda,t)}{\partial t}
=
k s(t)\big(\lambda-\bar\lambda(t)\big)x(\lambda,t)
+
\frac{\mathcal D}{2}
\frac{\partial^2 x(\lambda,t)}{\partial \lambda^2}.
\label{eq:SM_replicator}
\end{equation}
Equation~(\ref{eq:SM_replicator}) has the structure of a continuous
replicator--mutator equation: the selection term
$k s(t)(\lambda-\bar\lambda(t))x$ represents fitness-dependent growth
relative to the population mean, while the diffusion term
$(\mathcal D/2)\partial_{\lambda\lambda}x$ models mutation in trait space.
The effective fitness $f(\lambda,t)=k\lambda s(t)$ depends on the global
susceptible fraction $s(t)$, which acts as a time-dependent mean field
encoding resource depletion.

\subsection*{Discrete formulation}

Starting from the continuum trait dynamics,
\begin{equation}
\frac{\partial \rho_I(\lambda,t)}{\partial t}
=
k\,\lambda\,\rho_I(\lambda,t)\,s(t)
-\mu\,\rho_I(\lambda,t)
+\frac{\mathcal D}{2}\,
\frac{\partial^2 \rho_I(\lambda,t)}{\partial \lambda^2},
\label{eq:SM_continuum}
\end{equation}
we discretize trait space as $\lambda_i=\lambda_0+i\,\Delta\lambda$ and define
$\rho_i^I(t)\equiv \rho_I(\lambda_i,t)$. Using the centered finite-difference
approximation,
\begin{equation}
\left.\frac{\partial^2 \rho_I}{\partial \lambda^2}\right|_{\lambda=\lambda_i}
\simeq
\frac{\rho_{i-1}^I-2\rho_i^I+\rho_{i+1}^I}{(\Delta\lambda)^2},
\end{equation}
Eq.~(\ref{eq:SM_continuum}) becomes
\begin{equation}
\dot{\rho}_i^{I}
=
\lambda_i k \rho_i^{I}s
-\mu \rho_i^{I}
+\frac{\mathcal D}{2(\Delta\lambda)^2}\,
\left(\rho_{i-1}^I-2\rho_i^I+\rho_{i+1}^I\right).
\label{eq:SM_discrete_from_continuum_1}
\end{equation}
Introducing the discrete mutation operator $\Delta\rho_i^{I}
\equiv
\frac{\rho_{i-1}^{I}+\rho_{i+1}^{I}}{2}-\rho_i^{I}$,
we can rewrite Eq.~(\ref{eq:SM_discrete_from_continuum_1}) as
\begin{equation}
\dot{\rho}_i^{I}
=
\lambda_i k \rho_i^{I}s
-\mu \rho_i^{I}
+
D\,\Delta\rho_i^{I},
\qquad
D\equiv \frac{\mathcal D}{(\Delta\lambda)^2}.
\label{eq:SM_discrete_from_continuum_2}
\end{equation}
Therefore, changing the discretization $\Delta\lambda\to a\,\Delta\lambda$
can be exactly compensated by $D\to D/a^2$, leaving the continuum parameter
$\mathcal D=D(\Delta\lambda)^2$ invariant.

\clearpage
\newpage

\section*{Supplementary Note 2: Early-time dynamics}
In this Supplementary Note we derive the coupled evolution of the average
infectivity and the epidemic prevalence, making explicit the respective roles of
selection and mutation in shaping the early-time dynamics.
We start considering the discrete version of the model introduced in Eq.~(\ref{eq:SM_discrete_from_continuum_2}), where the density of infected
individuals carrying infectivity $\lambda_i$ evolves as
\begin{equation}
\dot{\rho}_i^{I}
=
\lambda_i k \rho_i^{I}s
-
\mu \rho_i^{I}
+
D\,\Delta\rho_i^{I},
\qquad
\Delta\rho_i^{I}
=
\frac{\rho_{i-1}^{I}+\rho_{i+1}^{I}}{2}-\rho_i^{I},
\label{eq:SM2_dynamics}
\end{equation}
with $s(t)=1-\sum_i \rho_i^{I}(t)-r(t)$.
We recall that the total prevalence and the average infectivity are defined as
\begin{equation}
i(t)\equiv \sum_i \rho_i^{I}(t),
\qquad
\bar{\lambda}(t)\equiv
\frac{\sum_i \lambda_i \rho_i^{I}(t)}{\sum_i \rho_i^{I}(t)} .
\label{eq:SM2_defs}
\end{equation}

\subsection*{Evolution of the average infectivity}

Let $A(t)\equiv \sum_i \lambda_i \rho_i^{I}(t)$ and $B(t)\equiv \sum_i \rho_i^{I}(t)=i(t)$,
so that $\bar{\lambda}=A/B$. Differentiating the average infectivity yields
\begin{equation}
\dot{\bar{\lambda}}
=
\frac{\dot A}{B}
-
\bar{\lambda}\,\frac{\dot B}{B}.
\label{eq:SM2_lambdabar_general}
\end{equation}
Using Eq.~\eqref{eq:SM2_dynamics},
\begin{align}
\dot A
&=
ks \sum_i \lambda_i^2 \rho_i^{I}
-\mu \sum_i \lambda_i \rho_i^{I}
+
D \sum_i \lambda_i \Delta\rho_i^{I},
\label{eq:SM2_A}\\
\dot B
&=
ks \sum_i \lambda_i \rho_i^{I}
-\mu \sum_i \rho_i^{I}.
\label{eq:SM2_B}
\end{align}
Substituting Eqs.~\eqref{eq:SM2_A}--\eqref{eq:SM2_B} into
Eq.~\eqref{eq:SM2_lambdabar_general}, the recovery terms cancel and we obtain
\begin{equation}
\dot{\bar{\lambda}}(t)
=
ks(t)\,
\mathrm{Var}(\lambda ;t)
+
D\,
\frac{\sum_i \lambda_i \Delta\rho_i^{I}(t)}{\sum_i \rho_i^{I}(t)},
\label{eq:SM2_lambdabar_split}
\end{equation}
where $\mathrm{Var}(\lambda ; t)
\equiv
\frac{\sum_i \lambda_i^2 \rho_i^{I}(t)}{\sum_i \rho_i^{I}(t)}
-
\bar{\lambda}(t)^2$
is the prevalence-weighted variance of infectivity. To lighten the notation, we will omit the time dependence in the variance in what follows. 

Equation~\eqref{eq:SM2_lambdabar_split} separates two contributions. The first term
is a \emph{selection} term: transmission favors larger $\lambda$, so the mean
infectivity increases proportionally to the width of the trait distribution,
modulated by the susceptible fraction $s(t)$. The second term is a
\emph{mutation} term, representing the net drift of the first moment induced by
diffusion in trait space. We note that the mutation term is a discrete divergence (the net difference of nearest-neighbor fluxes in trait space). Using $\lambda_{i+1}-\lambda_i=\Delta\lambda$,
it can be rewritten as
\begin{align}
\sum_i \lambda_i \Delta\rho_i^{I}
&=
\frac{1}{2}\sum_i \lambda_i
\left(\rho_{i+1}^{I}-2\rho_i^{I}+\rho_{i-1}^{I}\right)=-
\frac{\Delta\lambda}{2}
\left(\sum_i \rho_{i+1}^{I}-\sum_i \rho_{i-1}^{I}\right),
\label{eq:SM2_boundary_discrete}
\end{align}
which vanishes for symmetric trait domains with negligible density at the
boundaries (or, equivalently, under no-flux boundary conditions). Therefore, at
leading order,
\begin{equation}
\dot{\bar{\lambda}}(t)
\simeq
ks(t)\,\mathrm{Var}(\lambda).
\label{eq:SM2_lambdabar_sel}
\end{equation}

\subsubsection*{Diffusive growth of trait variance}

To close Eq.~\eqref{eq:SM2_lambdabar_sel} we determine the time evolution of
$\mathrm{Var}(\lambda)$.
Let $M_2(t)\equiv \sum_i \lambda_i^2 \rho_i^{I}(t)$, which corresponds to the second (non-central) moment of the infectivity distribution
weighted by the infected density.
Differentiating $\mathrm{Var}(\lambda)=M_2/B-\bar{\lambda}^2$ gives
\begin{equation}
\frac{d}{dt}\mathrm{Var}(\lambda)
=
\frac{\dot M_2}{B}
-
\frac{M_2}{B^2}\dot B
-
2\bar{\lambda}\dot{\bar{\lambda}}.
\label{eq:SM2_var_derivative_general}
\end{equation}

We first focus on the diffusive contribution to $\dot M_2$, which after recalling that $\sum_i\lambda_i\Delta\rho_i^I=0$ reads
\begin{equation}
\dot M_2\Big|_{\rm diff}
=
D\sum_i \lambda_i^2\,\Delta\rho_i^{I}.
\label{eq:SM2_M2_diff_discrete}
\end{equation}
In the continuum limit of Supplementary Note~1, $
D\,\Delta\rho^{I}
\;\to\;
\frac{\mathcal D}{2}\,
\frac{\partial^2\rho^{I}}{\partial \lambda^2}$, where $\mathcal D\equiv D(\Delta\lambda)^2$,
so that
\begin{equation}
\dot M_2\Big|_{\rm diff}
\simeq
\frac{\mathcal D}{2}
\int \lambda^2
\frac{\partial^2\rho^{I}}{\partial \lambda^2}
\,{\rm d}\lambda .
\label{eq:SM2_M2_diff_cont}
\end{equation}
Integrating by parts twice gives
\[
\int_{\lambda_{\min}}^{\lambda_{\max}}
\lambda^2 \partial_{\lambda\lambda}\rho^I\,d\lambda
=
\left[
\lambda^2\partial_\lambda\rho^I
-
2\lambda\rho^I
\right]_{\lambda_{\min}}^{\lambda_{\max}}
+
2\int_{\lambda_{\min}}^{\lambda_{\max}}
\rho^I(\lambda,t)\,d\lambda .
\]
In the early-time regime considered here, the trait distribution remains
localized far from the boundaries of the numerical trait domain. Hence
\(\rho^I\) and \(\partial_\lambda\rho^I\) are negligible at
\(\lambda_{\min}\) and \(\lambda_{\max}\), so the boundary term can be
neglected. Under this effective unbounded-domain approximation,
\begin{equation}
\int
\lambda^2 \partial_{\lambda\lambda}\rho^I\,d\lambda
\simeq
2\int \rho^I(\lambda,t)\,d\lambda .
\end{equation}
Since $\int \rho^{I}\,{\rm d}\lambda=B$, we obtain
\begin{equation}
\dot M_2\Big|_{\rm diff}
\simeq
\mathcal D\,B.
\label{eq:SM2_M2_diff_final}
\end{equation}
Diffusion conserves total prevalence, $\dot B|_{\rm diff}=0$, so the leading
diffusive contribution to the variance evolution is
\begin{equation}
\dot{\mathrm{Var}}(\lambda)\Big|_{\rm diff}
\simeq
\mathcal D.
\label{eq:SM2_var_dot_diff}
\end{equation}

The remaining contribution to $\dot{\mathrm{Var}}(\lambda)$ arises from
the selection terms in Eq.~\eqref{eq:SM2_dynamics}. After expressing
$\dot{\mathrm{Var}}(\lambda)= \dot M_2/B-(M_2/B^2)\dot B-2\bar{\lambda}\dot{\bar{\lambda}}$,
this contribution can be written as a central third-order moment of the trait
distribution, i.e. 
$\dot{\mathrm{Var}}(\lambda)\Big|_{\rm sel}=ks\langle(\lambda-\bar{\lambda})^3\rangle_{\rho^{I}}$.
The trait distribution forms a nearly symmetric
traveling profile in $\lambda$-space, so its skewness is small. Equivalently,
at leading order, the distribution is well described by a Gaussian
approximation, for which the third central moment vanishes. As a result, this
selection-induced contribution is subleading for the sharply peaked initial
condition considered here compared to the diffusive term. Therefore, integrating Eq. (\ref{eq:SM2_var_dot_diff}) yields
\begin{equation}
\mathrm{Var}(\lambda)
\simeq
\mathcal D\,t,
\label{eq:SM2_var_growth}
\end{equation}
where we used $\mathrm{Var}(\lambda; t=0)\simeq 0$ for a sharply peaked initial condition.

\subsubsection*{Closed expression for the evolution of the average infectivity}

Finally, substituting Eq.~\eqref{eq:SM2_var_growth} into
Eq.~\eqref{eq:SM2_lambdabar_sel} yields $\dot{\bar{\lambda}}(t)
\simeq
k\,\mathcal D\,s(t)\,t$, whose integration results in
\begin{equation}
\bar{\lambda}(t)
=
\lambda_0
+
k\mathcal D
\int_0^t s(t')\,t'\,{\rm d}t',
\label{eq:SM2_lambdabar_integral}
\end{equation}
as reported in the main text. At early times $s(t')\simeq 1$, giving Eq. (6):
\begin{equation}
\bar{\lambda}(t)\Big|_{t\to 0}
=
\lambda_0
+
\frac{1}{2}k\mathcal D\,t^2 .
\label{eq:SM2_lambdabar_early}
\end{equation}

\subsection*{Closed expression for the evolution of the prevalence}

Considering the early-time regime $s(t)\simeq 1$ and the evolution of the average infectivity given by
Eq.~\eqref{eq:SM2_lambdabar_early}, we can rewrite Eq.~\eqref{eq:SM_i_dot} as
\begin{equation}
\frac{\partial i(t)}{\partial t}\frac{1}{i(t)}
\simeq
k\lambda_0-\mu+\frac{1}{2}k^2\mathcal D\,t^2
=
\mu(R_0-1)
+
\frac{1}{2}k^2\mathcal D\,t^2 ,
\label{eq:SM2_i_logder}
\end{equation}
where $R_0\equiv k\lambda_0/\mu$.
Integrating Eq.~\eqref{eq:SM2_i_logder} from $0$ to $t$ finally yields Eq. (7):
\begin{equation}
i(t)\Big|_{t\to 0}
=
i_0
\exp\!\left[
\mu(R_0-1)t
+
\frac{k^2\mathcal D}{6}\,t^3
\right],
\label{eq:SM2_i_early}
\end{equation}
which makes explicit the superexponential early-time growth induced by infectivity
evolution.

\newpage

\section*{Supplementary Note 3: Additional features of the baseline epidemic dynamics}

{In this Supplementary Note, we further characterize two features of the baseline epidemic dynamics discussed in Fig.~1d of the main text: the apparent onset of the active phase at finite values of the initial reproduction number $R_0$, and the non-monotonic dependence of the epidemic peak $i_{\max}$ on $R_0$.}

\subsection*{Role of the extinction cutoff}

\begin{figure}[h]
    \centering
    \includegraphics[width=0.35\linewidth]{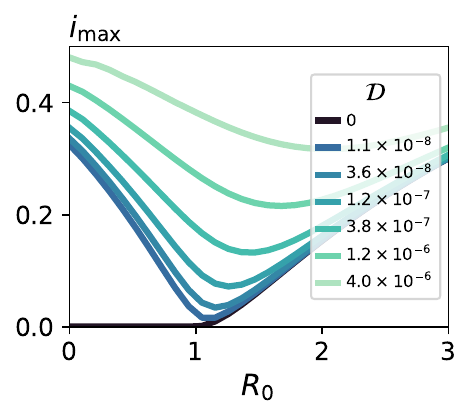}
\caption{\justifying{\textbf{No-cutoff version of the epidemic diagram shown in Fig.~1d.}
Epidemic peak $i_{\max}$ as a function of the initial basic reproduction number $R_0$, for different values of the effective diffusion rate $\mathcal{D}$. In contrast to Fig.~1d, the deterministic simulations are not stopped when $i(t)<10^{-4}$. We set $i_0=10^{-3}$, $\mu=1/7$, $k=10$, and $\Delta\lambda=10^{-3}$.}}
    \label{fig:cutoff}
\end{figure}

{In Fig.~1d of the main text, simulations are stopped whenever the prevalence falls below $i(t)=10^{-4}$. This cutoff mimics finite-population extinction, corresponding to fewer than one expected infected individual for populations of order $N\lesssim10^4$. To assess the effect of this cutoff, Supplementary Fig.~\ref{fig:cutoff} reproduces the epidemic diagram without cutoff. In this case, the active phase extends down to $R_0=0$ for any finite evolutionary rate, since even when the initial strain is non-transmissible, mutations can generate strains with $\lambda>0$.}

\subsection*{Origin of the non-monotonic dependence on $R_0$}

\begin{figure}[h]
    \centering
    \includegraphics[width=0.35\linewidth]{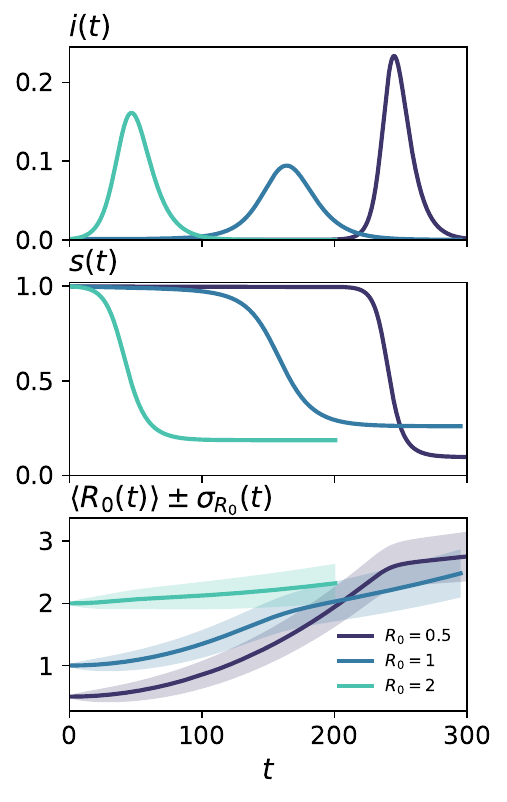}
\caption{\justifying{\textbf{Temporal dynamics of the epidemic and of the evolving infectivity distribution.}
\textbf{a} Time evolution of the prevalence $i(t)$ for three different initial values of $R_0$.
\textbf{b} Corresponding evolution of the susceptible fraction $s(t)$.
\textbf{c} Time evolution of the mean effective reproduction number $\langle R_0(t)\rangle$. Shaded regions indicate one standard deviation, $\langle R_0(t)\rangle\pm\sigma_{R_0}(t)$, computed from the infectivity distribution. Curves are shown until the prevalence falls below $10^{-4}$. We set $i_0=10^{-3}$, $\mu=1/7$, $k=10$, and $\mathcal{D}=10^{-7}$.}}
\label{fig:R0_time_evolution}
\end{figure}

{Supplementary Fig.~\ref{fig:R0_time_evolution} illustrates the dynamical origin of the non-monotonic dependence of $i_{\max}$ on the initial reproduction number. The key point is that the epidemic peak is not determined solely by the initial $R_0$, but also by the infectivity distribution reached when the epidemic takes off, while the susceptible population is still barely depleted. For $R_0=2$, the epidemic grows rapidly, leaving little time for infectivity evolution before take-off. For $R_0=1$, the slower growth allows the infectivity distribution to broaden further, although not enough to exceed the effective infectivity reached in the $R_0=2$ case. For $R_0=0.5$, instead, the pathogen remains at low prevalence for much longer. During this period, diffusion broadens the infectivity distribution and generates a high-$\lambda$ tail that is subsequently amplified by transmission. As a result, the epidemic eventually grows with a larger effective reproduction number and reaches a higher peak.}

\clearpage

\newpage

\section*{Supplementary Note 4: Hybrid approximation for epidemic control}

In this Supplementary Note we derive the hybrid analytical framework used to
characterize epidemic control in the presence of infectivity evolution. The
approach considers an intervention applied over a finite time window $[0,\tau]$,
during which infectivity evolves under control, followed by a post-intervention
phase in which the dynamics proceeds without control and with infectivity fixed
to its value at lifting.

\subsection*{Early-time dynamics under intervention}

We focus on control strategies applied during a finite intervention window
$[0,\tau]$, and distinguish between interventions acting on transmission
parameters, either by reducing the contact rate $k$ or the transmission
probability $\lambda$ by a factor $\varepsilon\in(0,1)$, and interventions acting
on the recovery rate $\mu$.
As shown in Supplementary Note~2, the evolution of the average infectivity follows from differentiating $\bar\lambda=\sum_i \lambda_i\rho_i^I/\sum_i\rho_i^I$ and substituting the dynamical equation for $\dot\rho_i^I$ (Eq. (\ref{eq:SM2_dynamics}), which is the discrete version of Eq. (1)). Importantly, only the transmission term $\lambda_i k \rho_i^I s$
contributes to the selective amplification of higher-infectivity strains, whereas the recovery term cancels out in the derivation of  $\dot{\bar{\lambda}}(t)$, yielding $\dot{\bar{\lambda}}(t)
=
k\,s(t)\,\mathrm{Var}(\lambda)$ (see Supplementary Note 2), which depends explicitly on the transmission process but not on the recovery
rate.

As a consequence, control interventions modify infectivity evolution only if
they rescale the transmission term in Eq.~(\ref{eq:SM2_dynamics}). For $k$ or $\lambda$ control,
the transmission term is multiplied by $\varepsilon$, so the same factor
appears in the selection term driving $\dot{\bar\lambda}$. In contrast,
$\mu$ control alters only the recovery term, which does not contribute to $\dot{\bar{\lambda}}(t)$, and therefore does not affect the selective
amplification mechanism.
Accordingly, the effect of control on infectivity evolution can be encoded in a
single parameter $\alpha$,  capturing whether the intervention rescales the
selection term. During the intervention, the evolution of the average infectivity
can be written in the unified form
\begin{equation}
\dot{\bar{\lambda}}(t)
=
\alpha\,k\,s(t)\,\mathrm{Var}(\lambda),
\label{eq:SM5_selection_alpha}
\end{equation}
where $\alpha=\varepsilon$ for interventions acting on transmission ($k$ or
$\lambda$ control), and $\alpha=1$ for interventions acting on the recovery rate
$\mu$.

Assuming that the intervention occurs during early-time dynamics, prevalence remains small and we approximate
$s(t)\simeq 1$. Using the early-time growth of the variance derived in Eq. (\ref{eq:SM2_var_growth}) of
Supplementary Note~2, $\mathrm{Var}(\lambda)\simeq\mathcal D\,t$,
integration of Eq.~\eqref{eq:SM5_selection_alpha} yields Eq. (8):
\begin{equation}
\bar{\lambda}(t,\varepsilon)
=
\lambda_0
+
\frac{1}{2}\,\alpha k\,\mathcal D\,t^2 .
\label{eq:SM5_lambdat_general}
\end{equation}

The evolution of the epidemic prevalence during an early-time intervention follows from
the equation
\begin{equation}
\frac{d i(t)}{d t}\frac{1}{i(t)}
=
\beta_c(t)-\mu_c ,
\label{eq:SM5_i_general}
\end{equation}
where $\beta_c(t)$ and $\mu_c$ denote the effective transmission and recovery
rates under control.

For interventions acting on transmission ($k$ or $\lambda$ control), the force
of infection is rescaled by a factor $\varepsilon$, yielding
$\beta_c(t)=\varepsilon k\,\bar{\lambda}(t)$ and $\mu_c=\mu$. In contrast, for interventions acting on the recovery rate ($\mu$ control), the
transmission process is unaltered, so that $\beta_c(t)=k\,\bar{\lambda}(t)$,
whereas the recovery rate is rescaled as $\mu_c=\mu/\varepsilon$.
Substituting
Eq.~\eqref{eq:SM5_lambdat_general} into Eq.~\eqref{eq:SM5_i_general} and integrating
from $0$ to $t$ gives
\begin{equation}
i(t,\varepsilon)|_{t\to 0}=
i_0
\exp\!\left[
\frac{\mu\alpha}{\varepsilon} \left(\varepsilon R_0-1 \right)t
+
\frac{\alpha^2 k^2}{6}\,
\mathcal D\,t^3
\right].
\label{eq:SM5_i_tau_general_}
\end{equation}
which coincides with Eq.~(9).
Eq.~\eqref{eq:SM5_i_tau_general_} shows explicitly the
coexistence of a linear epidemiological contribution, controlled by the
intervention, and a positive evolutionary contribution scaling as $t^3$.

\subsection*{Variation of the post-intervention peak}

We now derive an explicit expression for the variation of the post-intervention
epidemic peak with respect to the lifting time $\tau$ within the hybrid
approximation introduced above.
We assume that the global peak occurs after lifting the intervention, i.e., the
maximum prevalence is attained for $t>\tau$, and denote by
$(s(\tau),i(\tau),r(\tau))$ the state at the lifting time. After $t=\tau$, we
approximate the system dynamics to follow an uncontrolled SIR process with
recovery rate $\mu$ and transmission rate $\beta(\tau)$ set by the infectivity at
lifting, $\beta(\tau)=k\bar{\lambda}(\tau)$. The basic reproduction number immediately
after lifting is therefore $\mathcal R_0(\tau)=\beta(\tau)/\mu
=k\bar{\lambda}(\tau)/\mu$.

For an SIR outbreak with constant basic reproduction number $R_0$ starting from
$(s_0,i_0,r_0)$, the infected fraction reaches its maximum when $s=1/R_0$.
Using the SIR invariant
\begin{equation}
r(t)+\frac{1}{R_0}\ln s(t)
=
r_0+\frac{1}{R_0}\ln s_0,
\label{eq:SM5_invariant}
\end{equation}
and imposing the condition $s_{\rm peak}=1/R_0$, one obtains the peak
prevalence
\begin{equation}
i_{\max}
=
1-r_0-\frac{1}{R_0}
\left[1+\ln(R_0 s_0)\right].
\label{eq:SM5_imax_general}
\end{equation}

Applying Eq.~\eqref{eq:SM5_imax_general} to the post-lifting outbreak with
$(s_0,i_0,r_0)=(s(\tau),i(\tau),r(\tau))$ and $R_0=\mathcal R_0(\tau)$ yields
Eq.~(B1):
\begin{equation}
i_{\max}(\tau)
=
1-r(\tau)
-\frac{1}{\mathcal R_0(\tau)}
\left[1+\ln\!\big(\mathcal R_0(\tau)s(\tau)\big)\right],
\label{eq:SM5_imax_tau}
\end{equation}
where $s(\tau)=1-i(\tau)-r(\tau)$.

We now differentiate Eq.~\eqref{eq:SM5_imax_tau} with respect to $\tau$ using the
chain rule, which yields
\begin{align}
\frac{\mathrm{d} i_{\max}}{\mathrm{d}\tau}
&=
-\dot r(\tau)
+
\frac{\dot{\mathcal R_0}(\tau)}{\mathcal R_0(\tau)^2}
\left[1+\ln\!\big(\mathcal R_0(\tau)s(\tau)\big)\right]
-\frac{1}{\mathcal R_0(\tau)}
\left(
\frac{\dot{\mathcal R_0}(\tau)}{\mathcal R_0(\tau)}
+\frac{\dot s(\tau)}{s(\tau)}
\right)\\
&=
-\dot r(\tau)
+
\frac{\dot{\mathcal R_0}(\tau)}{\mathcal R_0(\tau)^2}
\ln\!\big(\mathcal R_0(\tau)s(\tau)\big)
-
\frac{1}{\mathcal R_0(\tau)}
\frac{\dot s(\tau)}{s(\tau)}.
\label{eq:SM5_dimax_compact}
\end{align}

During the controlled phase $t\in[0,\tau]$, the SIR equations imply
\begin{equation}
\dot r(\tau)=\mu_c\,i(\tau),
\qquad
\dot s(\tau)=-\beta_c(\tau)\,s(\tau)\,i(\tau),
\label{eq:SM5_sr_dots_general}
\end{equation}
where $\beta_c(\tau)$ and $\mu_c$ denote the controlled transmission and recovery
rates, respectively. Recall that the post-lifting outbreak is approximated as an
uncontrolled SIR process with transmission rate $\beta(\tau)=k\bar{\lambda}(\tau)$
and baseline recovery rate $\mu$.
Using $\dot s(\tau)/s(\tau)=-\beta_c(\tau)\,i(\tau)$, Eq.~\eqref{eq:SM5_dimax_compact}
can be rewritten as
\begin{equation}
\frac{\mathrm{d} i_{\max}}{\mathrm{d}\tau}
=
-\mu_c\,i(\tau)
+
\frac{\dot{\mathcal R_0}(\tau)}{\mathcal R_0(\tau)^2}\,
\ln\!\big(\mathcal R_0(\tau)s(\tau)\big)
+
\frac{\beta_c(\tau)}{\mathcal R_0(\tau)}\,i(\tau).
\label{eq:SM5_dimax_general_rates}
\end{equation}
Introducing $\alpha$ such that $\alpha=\varepsilon$ for
interventions acting on transmission ($k$ or $\lambda$ control),
whereas $\alpha=1$ for interventions
acting on the recovery rate ($\mu$ control), Eq.~(\ref{eq:SM5_dimax_general_rates}) can be written
in the compact form
\begin{equation}
\frac{d i_{\max}}{d\tau}
=
-\mu\, i(\tau)\,\alpha\frac{(1-\varepsilon)}{\varepsilon}
+
\frac{\dot{\mathcal R_0} (\tau)}{\mathcal R_0(\tau)^2}
\ln\!\big(\mathcal R_0(\tau)s(\tau)\big).
\label{eq:S_41}
\end{equation}
Finally, using the early-time evolution of the average infectivity
during the intervention, $\bar\lambda(\tau)=\lambda_0+\alpha k D\tau^2/2$,
we obtain
\begin{equation}
\dot{\mathcal R_0}(\tau)
=
\frac{\alpha k^2}{\mu} D \tau,
\label{eq:S_43}
\end{equation}
which, substituted into Eq.~(\ref{eq:S_41}), yields the unified expression reported in Eq.~(10) of the main text.

\subsection*{Critical diffusion strength}

The onset of nonmonotonic epidemic responses is determined by the existence of
intervention durations $\tau$ for which the post-lifting epidemic peak satisfies
$\mathrm{d} i_{\max}/\mathrm{d}\tau=0$.
Using Eq.~(10) of the main text and evaluating all quantities at lifting, this condition
can be written explicitly as
\begin{equation}
\mu\,\alpha\frac{(1-\varepsilon)}{\varepsilon}\,i(\tau)
=
\frac{\alpha k^2}{\mu}\,
\mathcal D\,\tau\,
\frac{\ln\!\big(\mathcal R_0(\tau)s(\tau)\big)}{\mathcal R_0(\tau)^2},
\label{eq:SM5_balance_general}
\end{equation}
where $\alpha=\varepsilon$ for transmission-based control ($k$ or $\lambda$ control) and
$\alpha=1$ for recovery-based control ($\mu$ control).
Equation~\eqref{eq:SM5_balance_general} expresses a balance between a negative
epidemiological contribution proportional to the prevalence accumulated during the
intervention (left-hand side) and a positive evolutionary contribution driven by the
increase of the time-varying basic reproduction number due to infectivity evolution (right-hand side).
Since $\alpha>0$, it cancels from both sides, yielding the implicit relation reported in Eq. (C1):
\begin{equation}
\mathcal D_c(\tau)
=
\frac{\mu^2(1-\varepsilon)\,i(\tau)\,\mathcal R_0(\tau)^2}
{\varepsilon k^2\,\tau\,\ln \!\big(\mathcal R_0(\tau)s(\tau)\big)}.
\label{eq:SM5_Dc_tau_general}
\end{equation}

To obtain an explicit expression, we assume that the prevalence remains small at
lifting and that the dominant contribution encoding evolutionary effects enters through
the prevalence $i(\tau)$. Accordingly, we approximate $s(\tau)\simeq 1$,
$\ln\!\big(\mathcal R_0(\tau)s(\tau)\big)\simeq \ln R_0$ and
$\mathcal{R}_0(\tau)\simeq R_0$.
Under these approximations, Eq.~\eqref{eq:SM5_Dc_tau_general} reduces to
\begin{equation}
\mathcal D_c(\tau)\simeq
\frac{\mu^2(1-\varepsilon)\,R_0^2}{\varepsilon k^2\,\tau\,\ln R_0}\;
i(\tau).
\label{eq:SM5_Dc_tau_early}
\end{equation}

We now substitute the early-time prevalence at lifting in its unified form,
\begin{equation}
i(\tau)\simeq
i_0
\exp\!\left[
\frac{\mu\alpha}{\varepsilon}\left(\varepsilon R_0-1\right)\tau
+
\frac{\alpha^2 k^2}{6}\,
\mathcal D_c(\tau)\,\tau^3
\right],
\label{eq:SM5_i_tau_for_Dc_general}
\end{equation}
where $\alpha=\varepsilon$ for $k$  or $\lambda$ control and
$\alpha=1$ for $\mu$ control.
Substituting Eq.~\eqref{eq:SM5_i_tau_for_Dc_general} into Eq.~\eqref{eq:SM5_Dc_tau_early} yields
\begin{equation}
\mathcal D_c(\tau)\simeq
\frac{\mu^2(1-\varepsilon)\,R_0^2}{\varepsilon k^2\,\tau\,\ln R_0}\;
i_0
\exp\!\left[
\frac{\mu\alpha}{\varepsilon}\left(\varepsilon R_0-1\right)\tau
+
\frac{\alpha^2 k^2}{6}\,
\mathcal D_c(\tau)\,\tau^3
\right].
\label{eq:SM5_Dc_simplified_general}
\end{equation}

Defining
\begin{equation}
P(\tau)=\frac{1}{6}\,\alpha^2 k^2\,\tau^3,
\qquad
Q(\tau)=
\frac{\mu^2(1-\varepsilon)\,R_0^2}{\varepsilon k^2\,\tau\,\ln R_0}\;
i_0\,
\exp\!\left[\frac{\mu\alpha}{\varepsilon}\left(\varepsilon R_0-1\right)\tau\right],
\label{eq:SM5_Bc_def_general}
\end{equation}
Eq.~\eqref{eq:SM5_Dc_simplified_general} can be written as
$-P(\tau)\mathcal D_c(\tau)\,e^{-P(\tau)\mathcal D_c(\tau)}=-P(\tau)Q(\tau)$,
so that the solution can be expressed in closed form as in Eq. (C2) using the Lambert $W$ function,
defined by $W(x)e^{W(x)}=x$:
\begin{equation}
\mathcal D_c(\tau)=
-\frac{1}{P(\tau)}\,
W_0\!\left[-P(\tau)Q(\tau)\right].
\label{eq:SM5_Dc_lambert_general}
\end{equation}

Here $W_0$ denotes the relevant real branch of the Lambert function. Due to the assumptions
of our hybrid approximation, this expression diverges $\mathcal D_c(\tau)\to\infty$ as
$\tau\to0$, while capturing the existence of a finite critical effective diffusion strength
for small but finite intervention durations.

\subsection*{Critical intervention duration}

When evolutionary effects are present, the post-lifting epidemic peak
$i_{\max}(\tau)$ is no longer monotonic in the intervention duration $\tau$.
An interior extremum $\tau^\star>0$ is determined by the condition
$\mathrm{d} i_{\max}/\mathrm{d}\tau=0$.
Using Eq. (8) of the main text and evaluating all quantities at the lifting
time, the condition $\mathrm{d} i_{\max}/\mathrm{d}\tau=0$ can be written as
\begin{equation}
\mu\,\frac{(1-\varepsilon)}{\varepsilon}\, i(\tau)
=
\frac{k^2}{\mu}\,
\mathcal D\,\tau\,
\frac{\ln\!\big(\mathcal R_0(\tau)s(\tau)\big)}{\mathcal R_0(\tau)^2},
\label{eq:SM5_tau_star_balance_general}
\end{equation}
Assuming early-time dynamics and that the dominant evolutionary contribution
enters through the prevalence at lifting, we approximate
$s(\tau)\simeq 1$, $\mathcal R_0(\tau)\simeq R_0$, and
$\ln\!\big(\mathcal R_0(\tau)s(\tau)\big)\simeq \ln R_0$.
Under these conditions, Eq.~\eqref{eq:SM5_tau_star_balance_general} reduces to
\begin{equation}
\mu\,\frac{(1-\varepsilon)}{\varepsilon}\, i(\tau)
\simeq
\frac{k^2}{\mu}\,
\mathcal D\,\tau\,
\frac{\ln R_0}{ R_0^2}.
\label{eq:SM5_tau_star_balance_simplified}
\end{equation}
Substituting Eq.~(9) into
Eq.~\eqref{eq:SM5_tau_star_balance_simplified} yields the transcendental equation
\begin{equation}
C\,\mathcal D\,\tau
=
\mu\,\frac{(1-\varepsilon)}{\varepsilon}\, i_0
\exp\!\left[
A\tau
+
B\,\mathcal D\,\tau^3
\right],
\label{eq:SM5_tau_star_trans_general}
\end{equation}
with
\begin{equation}
A=\frac{\mu\alpha}{\varepsilon}\left(\varepsilon R_0-1\right),
\qquad
B=\frac{\alpha^2 k^2}{6},
\qquad
C=\frac{k^2}{\mu}\frac{\ln R_0}{ R_0^2},
\label{eq:SM5_ABC_general}
\end{equation}
and $\alpha$ encoding the effect of the control strategy, being $\alpha=\varepsilon$ for interventions acting on transmission ($k$ or $\lambda$ control), whereas $\alpha=1$ for interventions acting on the recovery rate ($\mu$ control).
In the regime where the cubic evolutionary term dominates the exponent,
$B\mathcal D\tau^3\gg |A\tau|$, Eq.~\eqref{eq:SM5_tau_star_trans_general} simplifies to
\begin{equation}
C\,\mathcal D\,\tau
\simeq
\mu\,\frac{(1-\varepsilon)}{\varepsilon}\, i_0\,
\exp\!\left(B\,\mathcal D\,\tau^3\right).
\label{eq:SM5_tau_star_cubic_general}
\end{equation}
Defining $x\equiv -3B\mathcal D\,\tau^3$, one obtains
\begin{equation}
x\,e^{x}
=
-\frac{3B\,\mu^3(1-\varepsilon)^3\,i_0^3}{\varepsilon^3C^3\,\mathcal D^2},
\label{eq:SM5_xeqx_general}
\end{equation}
and finally, using the definition of the Lambert function, $W(z)e^{W(z)}=z$, we reach Eq. (11) in the main text:
\begin{equation}
\tau^\star
\simeq
\left[
-\frac{1}{3B\mathcal D}\,
W_{-1}\!\left(
-\frac{3B\,\mu^3(1-\varepsilon)^3\,i_0^3}
{\varepsilon^3C^3\,\mathcal D^2}
\right)
\right]^{1/3}.
\label{eq:SM5_tau_star_lambert_general}
\end{equation}

The argument of the Lambert function is negative, so two real branches may exist
whenever $-1/e<z<0$. The physically relevant solution corresponding to an interior
maximum of $i_{\max}(\tau)$ is obtained from the $W_{-1}$ branch. The principal
branch $W_0$ yields a small-$\tau$ solution that corrects an initial decreasing
trend introduced by the hybrid approximation.

\newpage

\section{Supplementary Note 5: Final epidemic size under control}

In the main text, we focused on the epidemic peak \(i_{\max}\), since this quantity measures the maximum pressure exerted on the healthcare system during an outbreak. However, another epidemic observable is the final epidemic size, or attack rate
\begin{equation}
    r^\infty=\lim_{t\to\infty} r(t),
\end{equation}
which measures the total fraction of the population infected over the course of the epidemic.

In Supplementary Fig.~\ref{fig:control_temporal_cuts}, we compare the effect of \(\lambda\) control on both observables. Supplementary Fig.~\ref{fig:control_temporal_cuts}a shows the epidemic peak \(i_{\max}\) as a function of the intervention duration \(\tau\) (reproducing Fig. 2a of the main text), while Supplementary Fig.~\ref{fig:control_temporal_cuts}b shows the corresponding attack rate \(r^\infty\). The nonmonotonic response to the intervention duration is visible in both quantities, showing that the effect of infectivity evolution is consistent across short-term control interventions.

Nevertheless, \(i_{\max}\) and \(r^\infty\) do not always provide equivalent assessments of control efficiency. This distinction is especially clear for long interventions. As shown by comparing Supplementary Figs.~\ref{fig:control_temporal_cuts}c and d, increasing the intervention duration can strongly reduce the first epidemic wave, but it can also leave a low-prevalence infected population that resurges after the intervention is lifted. In such cases, the final size \(r^\infty\) can become large even when the control policy has reduced or delayed the main epidemic peak. This results in the abrupt shifts on $r^{\infty}$ reported in Supplementary Fig.~\ref{fig:control_temporal_cuts}b.

This difference reflects the fact that \(i_{\max}\) and \(r^\infty\) quantify distinct aspects of epidemic dynamics. The peak \(i_{\max}\) is controlled by the largest instantaneous burden during the outbreak, whereas \(r^\infty\) accumulates infections over the full trajectory, including delayed post-control resurgence. Therefore, in the presence of infectivity evolution, the attack rate is informative but might not be the best observable to assess control efficiency.

\begin{figure}[h]
    \centering
    \includegraphics[width=0.8\linewidth]{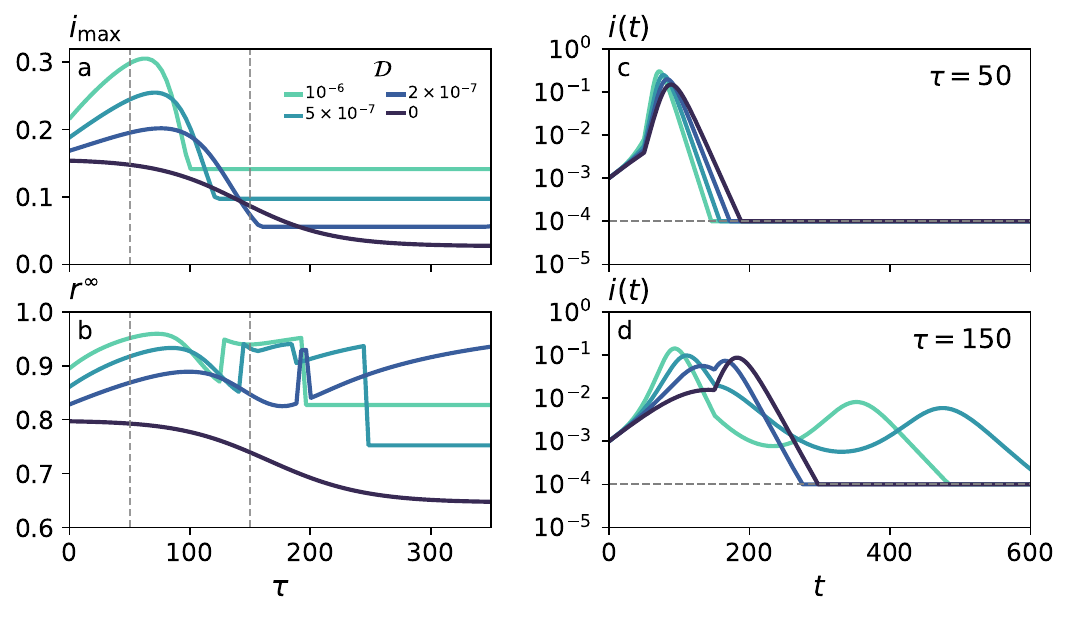}
    \caption{\justifying
\textbf{Effect of infectivity evolution on epidemic control.} 
\textbf{a} Epidemic peak \(i_{\max}\) under \(\lambda\) control as a function of the intervention time \(\tau\), for different values of the effective diffusion rate \(\mathcal{D}\). 
\textbf{b} Attack rate \(r^\infty\) as a function of the intervention time \(\tau\), for the same values of \(\mathcal{D}\). 
\textbf{c,d} Temporal evolution of \(i(t)\) for the two intervention durations marked by the vertical dashed lines in panels \textbf{a,b}: \textbf{c} \(\tau=50\) and \textbf{d} \(\tau=150\). The horizontal dashed line marks the threshold \(i(t)=10^{-4}\). 
In all panels, set \(i_0=10^{-3}\), \(R_0=2\), \(\mu=1/7\), \(k=10\), and \(\varepsilon=0.6\).}
    \label{fig:control_temporal_cuts}
\end{figure}

\section{Supplementary Note 6: Robustness of the nonmonotonic epidemic response}

In Fig.~2 of the main text, we showed that infectivity evolution
induces a nonmonotonic dependence of the post-intervention epidemic peak
$i_{\max}$ on the intervention duration $\tau$.
In Supplementary Fig. \ref{fig:SM_Fig1} we assess the robustness of this behavior under variations
of key epidemiological parameters, while keeping the effective diffusion
strength fixed at $\mathcal{D}=10^{-6}$.
We independently vary: the initial basic reproduction number $R_0$ in Supplementary Fig. \ref{fig:SM_Fig1}.a,
the intervention strength $\varepsilon$  in Supplementary Fig. \ref{fig:SM_Fig1}.b,
the activation threshold $i_{\mathrm{thr}}$  in Supplementary Fig. \ref{fig:SM_Fig1}.c,
the mutational bias \(p_-\) in Supplementary Fig.~\ref{fig:SM_Fig1}d,
and the policy starting time \(t_{\mathrm{start}}\) in Supplementary Fig.~\ref{fig:SM_Fig1}e.

For the activation-threshold protocol, the control policy is implemented once the infected
fraction satisfies \(i(t)=i_{\mathrm{thr}}\).
Thus, \(i_{\mathrm{thr}}=i_0\) corresponds to an intervention applied
immediately at the onset of the outbreak, whereas larger values
of \(i_{\mathrm{thr}}\) delay the implementation of control,
allowing the epidemic and infectivity evolution to progress
before intervention.
For the biased-mutation case, we modify the mutation operator in Eq.~(S.8) by assigning
different probabilities to mutations towards lower and higher infectivity values. In particular, the mutation term becomes
\[\Delta\rho_i^I\rightarrow
p_+\rho_{i-1}
+
p_-\rho_{i+1}
-
\rho_i,
\]
with \(p_++p_-=1\). Here \(p_-\) is the probability of mutating towards smaller
\(\lambda\), while \(p_+\) is the probability of mutating towards larger \(\lambda\).
The symmetric diffusive case corresponds to \(p_-=p_+=1/2\), whereas \(p_->1/2\)
introduces a negative drift in infectivity space.
Finally, for the delayed-start protocol, the intervention is applied at a prescribed time
\(t_{\mathrm{start}}\) and lasts for a duration \(\tau\).

In all cases, the qualitative nonmonotonic behavior persists. Increasing $\tau$ increases the post-lifting epidemic peak until a critical value $\tau^\star$. After that, the peak decreases with $\tau$.
Quantitatively, increasing $R_0$ shifts the maximum toward smaller
intervention durations.
Varying the intervention strength $\varepsilon$
modulates the balance between epidemiological suppression
and evolutionary amplification:
stronger interventions (smaller $\varepsilon$) advance the onset
of the maximum and reduce its amplitude.
Changes in $i_{\text{thr}}$ effectively rescale the initial conditions of the epidemic trajectories. This affects the early-time
accumulation of evolutionary effects and therefore alters
the location of the maximum. The biased-mutation results show that the mechanism is
robust to moderate negative drift, although strong bias towards lower infectivity
can reduce the amplitude of the effect by limiting the growth of the
high-infectivity tail. Similarly, delaying the start of the intervention affects dynamics in an equivalent way to the $i_{\text{thr}}$ mechanism.
Overall, these results confirm that the nonmonotonous behavior is robust across
a broad region of parameter space.

\begin{figure}[h!]
    \centering
    \includegraphics[width=1\linewidth]{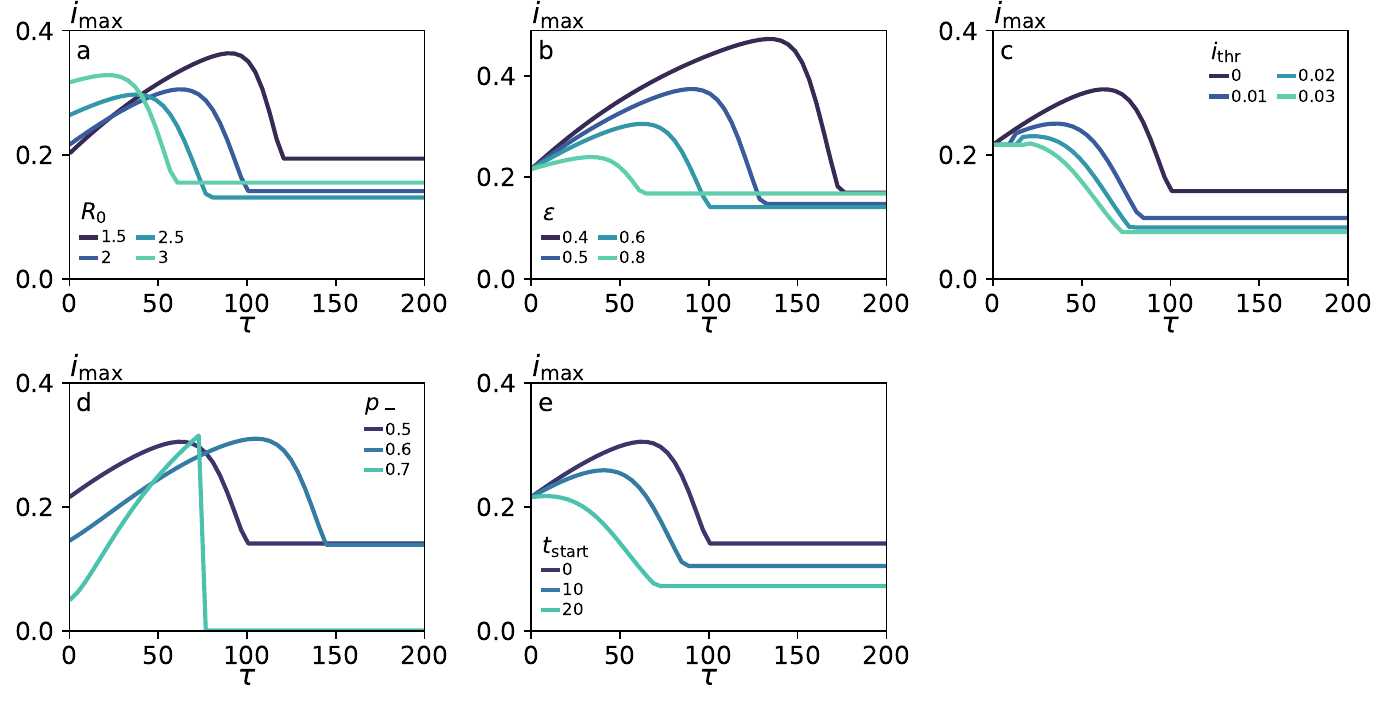}
    \caption{\justifying
    \textbf{Robustness of the nonmonotonic epidemic impact.}
    Epidemic peak $i_{\max}$ as a function of the intervention duration $\tau$.
    Panels show the dependence on:
    \textbf{a} the initial basic reproduction number $R_0$,
    \textbf{b} the intervention strength $\varepsilon$,
    \textbf{c} the activation threshold $i_{\text{thr}}$,
    \textbf{d} the mutational bias \(p_-\), and
    \textbf{e} the policy starting time \(t_{\mathrm{start}}\).
    Unless otherwise varied, parameters are
    $R_0=2$, $\varepsilon=0.6$, $i_{\text{thr}}=0$, $p_-=0.5$, $t_{\mathrm{start}}=0$,
    $i_0=10^{-3}$, $\mu=1/7$, $k=10$, $\mathcal{D}=10^{-6}$ and $\Delta\lambda=10^{-3}$.}
    \label{fig:SM_Fig1}
\end{figure}

\clearpage

\newpage

\clearpage

\newpage

\section*{Supplementary Note 7: Sigmoid approximation of susceptible depletion}

In this Supplementary Note we expand on Appendix~A by providing the detailed
derivation of the analytical approximation for the time dependence of the
susceptible fraction, and use it to characterize the
slowdown of infectivity growth at intermediate times.

In standard SIR dynamics, the susceptible population decreases monotonically from
$s(0)\simeq 1$ to its final value $s(\infty)=1-r(\infty)$. Although no closed-form expression exists for
$s(t)$, its qualitative shape is robust: a slow initial decay, a rapid drop around
the epidemic peak, and a final saturation.
Motivated by this behavior, we approximate the susceptible fraction by a sigmoid
centered at the epidemic peak time $t_p$,
\begin{equation}
s(t)
\simeq
s(\infty)
+
\frac{1-s(\infty)}{1+e^{\sigma(t-t_p)}}.
\label{eq:SM3_sigmoid}
\end{equation}
The parameters $s(\infty)$,
$t_p$, and $\sigma$ encode the final epidemic size, the time of maximum prevalence,
and the characteristic timescale of susceptible depletion, respectively.

Substituting the sigmoidal approximation~\eqref{eq:SM3_sigmoid} into Eq.\eqref{eq:SM2_lambdabar_integral}, which captures the evolution of the average infectivity, yields
\begin{equation}
\bar{\lambda}(t)
=
\lambda_0
+
k\mathcal{D}
\left[
s(\infty)\frac{t^2}{2}
+
\big(1-s(\infty)\big)
\int_0^t
\frac{t'}{1+e^{\sigma(t'-t_p)}}
\,{\rm d}t'
\right].
\label{eq:SM3_lambda_integral}
\end{equation}
To obtain a closed expression, we define the auxiliary integral $
\mathcal{J}(t)
\equiv
\int_0^t
\frac{t'}{1+e^{\sigma(t'-t_p)}}\,dt'$.
For times around and after the epidemic peak, we are interested in the
regime in which the logistic function has entered its post-peak tail. In
particular, for $\sigma(t-t_p)\gtrsim 1$,
we write
\[
\mathcal{J}(t)
=
\mathcal{J}_{\infty}
-
\int_t^\infty
\frac{t'}{1+e^{\sigma(t'-t_p)}}\,dt',\quad \quad\text{where}\quad
\mathcal{J}_{\infty}
=
\int_0^\infty
\frac{t'}{1+e^{\sigma(t'-t_p)}}\,dt'
\]
is a constant. Since the remaining integral only involves the post-peak
tail, we approximate $\frac{1}{1+e^{\sigma(t'-t_p)}}
\simeq
e^{-\sigma(t'-t_p)}$, provided $t'\geq t,\ \sigma(t-t_p)\gtrsim 1$.
Thus, $\mathcal{J}(t)
\simeq
\mathcal{J}_{\infty}
-
\int_t^\infty
t' e^{-\sigma(t'-t_p)}\,dt'$.
Performing the last integral and using \(t=t_p+(t-t_p)\), we obtain that
\begin{equation}
\mathcal{J}(t)
\simeq
\mathcal{J}_{\infty}
-
\left[
\frac{t-t_p}{\sigma}
+
\frac{t_p}{\sigma}
+
\frac{1}{\sigma^2}
\right]
e^{-\sigma(t-t_p)} ,
\label{eq:SM3_integral_eval}
\end{equation}
up to corrections of order \(e^{-2\sigma(t-t_p)}\). Substituting this
expression into Eq.~(\ref{eq:SM3_lambda_integral}), the constant contribution proportional to
\(\mathcal{J}_{\infty}\) can be absorbed into an additive offset of
\(\bar{\lambda}(t)\). Keeping the explicitly time-dependent transient
term that captures the post-peak slowdown, we obtain
\begin{equation}
\bar{\lambda}(t)
\sim
k\mathcal D
\left[
s(\infty)\frac{t^2}{2}
-
\left(1-s(\infty)\right)
\frac{\sigma t+1}{\sigma^2}
e^{-\sigma(t-t_p)}
\right],
\label{eq:SM3_lambda_sigmoid}
\end{equation}
which displays the negative transient contribution induced by
susceptible depletion around and after the epidemic peak.

\newpage

\end{document}